\documentclass[12pt,onecolumn,floatfix,altaffilletter,superscriptaddress, tightenlines,showpacs,showkeys,preprintnumbers,nofootinbib]{revtex4-1}
\pdfoutput=1
\usepackage[colorlinks=true,citecolor=blue,linkcolor=blue,breaklinks=true]{hyperref}
\usepackage{amsmath,amssymb}
\usepackage{mathtools}
\usepackage{epsfig} 
\usepackage{graphicx}        
\usepackage{url}
\usepackage{color}
\usepackage{multirow}
\usepackage{placeins}
\usepackage[dvipsnames]{xcolor}
\usepackage{braket}
\usepackage{float}
\usepackage{slashed}
\usepackage{tikz}
\hypersetup{
  colorlinks=true,
  linkcolor=red, 
  filecolor=magenta,   
  urlcolor=blue,
  citecolor=blue,
}

\allowdisplaybreaks

\bibpunct{[}{]}{,}{n}{}{,}

\newcommand{\dd}{\mathrm{d}}

\def\beq{\begin{equation}}
\def\eeq{\end{equation}}
\def\bea{\begin{eqnarray}}
\def\eea{\end{eqnarray}}
\makeatletter
\@addtoreset{equation}{section}

\begin{document}

\title{Testing Super-Heavy Dark Matter from Primordial Black Holes with Gravitational Waves}

\author{Rome Samanta}
\email{romesamanta@gmail.com}
\author{Federico R. Urban}
\email{federico.urban@fzu.cz}
\affiliation{CEICO, Institute of Physics of the Czech Academy of Sciences, Na Slovance 1999/2, 182 21 Prague 8, Czech Republic}

\begin{abstract} 
Ultra-light primordial black holes with masses $M_{BH}<10^9$~g evaporate before big-bang nucleosynthesis producing all matter fields, including dark matter, in particular super-heavy dark matter: $M_{DM}\gtrsim 10^{10}$~GeV.  If the dark matter gets its mass via $U(1)$ symmetry-breaking, the phase transition that gives a mass to the dark matter also produces cosmic strings which radiate gravitational waves.  Because the symmetry-breaking scale $\Lambda_{CS}$ is of the same order as $M_{DM}$, the gravitational waves radiated by the cosmic strings have a large enough amplitude to be detectable across all frequencies accessible with current and planned experimental facilities.  Moreover, an epoch of early primordial black hole domination introduces a unique spectral break in the gravitational wave spectrum whose frequency is related to the super-heavy dark matter mass.  Hence, the features of a stochastic background of primordial gravitational waves could indicate that super-heavy dark matter originated from primordial black holes.  In this perspective, the recent finding of a stochastic common-spectrum process across many pulsars by two nano-frequency pulsar timing arrays would fix the dark matter mass to be $3\times 10^{13}~\text{GeV} \lesssim M_{DM} \lesssim 10^{14}~\text{GeV}$.  The (non-)detection of a spectral break at $0.2~\text{Hz} \lesssim f_* \lesssim 0.4~\text{Hz}$ would (exclude) substantiate this interpretation of the signal.\\

\end{abstract} 

\maketitle

\section{introduction}

Ultra-light primordial black holes (PBHs), that is, black holes with mass $M_{BH}\lesssim 10^9$~g)~\cite{bh1,bh2,bh3} that exist and evaporate prior to Big Bang Nucleosynthesis (BBN) via Hawking radiation~\cite{hr}, can leave observable imprints by producing Gravitational Waves (GWs)~\cite{bhgw1,bhgw2,bhgw3,bhgw4,bhgw4a,bhgw5,bhgw6,bhgw7}, generating the Baryon asymmetry of the Universe (BAU)~\cite{br0,br0a,dm0,br1,br2,br3,br4,Aliferis:2020dxr} and cosmologically stable relics~\cite{khlo1,khlo2}.  Most importantly, owing to the fact that PBHs must be agnostic about Standard Model (SM) quantum numbers, as they evaporate they must also produce Dark Matter (DM)~\cite{dm0,dm1,dm2,dm3,dm4,dm4a,dm5,dm5a,dm6,dm7,dm10,dm11,Kitabayashi:2021hox,newdm1,newdm2,newdm3,newdm4,newdm5,newdm6}.

The existence of DM in the Universe is strongly corroborated by experimental data ~\cite{Planck:2015fie,Bull:2015stt}.  However, thus far only the gravitational effects of DM have been observed, and, despite extensive dedicated experimental programmes, any attempt to detect possible DM interactions with the SM have yielded null results\footnote{In fact, this is a good reason not to discount the possibility that what looks like a new particle is instead a deviation from General Relativity ~\cite{mond1,mond2}.}.  Therefore, it might well be that DM does not interact with the SM \emph{at all}.  In this case, we need to rethink about (a) how DM could be generated in the early Universe, and (b) how to further constrain and hopefully detect and test DM via its gravitational effects---possibly the only way to do so.  In this article we show how, if ultra-light PBHs existed in the early Universe, (a) DM, in fact super-heavy dark matter (SHDM) with masses above $M_{DM}\gtrsim10^{10}$~GeV, can be easily generated, and (b) the PBH origin of SHDM is testable through its peculiar features in the spectrum of primordial Gravitational Waves that this mechanism would produce.

Owing primarily to its very large energy scale, SHDM is very difficult to make. Existing mechanisms are, e.g., gravitational production at the end of inflation~\cite{shdm1,shdm2,shdm3,shdm4}, supersymmetry breaking~\cite{shdmstr},  freeze-in~\cite{shdmfi1,shdmfi2}, thermal freeze-out~\cite{shdmfo1,Kim:2019udq,shdmfo2,shdmfo3} and  phase transitions~\cite{Azatov:2021ifm,Ahmadvand:2021vxs}.  The mechanism of production of SHDM via PBHs, which we adopt in this work, has two distinct advantages.  First, with PBHs one can access a large range of SHDM masses, potentially up to the Planck scale, without having to specify the details of the couplings of the dark sector to the SM or gravity~\cite{sab1,sab2}.  Second, for a large enough initial PBH energy density,  PBHs come to dominate the energy density of the Universe before evaporation; when they finally evaporate they inject entropy in the system and significantly dilute any preexisting model-dependent SHDM abundance, thereby making the ${\rm PBH \rightarrow SHDM}$ channel unambiguously dominant.

SHDM is also difficult to test because it is too heavy to be produced directly in colliders, even assuming that SHDM interacts with the SM in the first place.  If SHDM does interact, it is expected to decay into ultra-high-energy cosmic rays (UHECRs), either charged nuclei or neutral photons or neutrinos~\cite{Kalashev:2016cre}.  The flux of UHECRs observed by experiments is very low, and there are large uncertainties about the energy spectrum and chemical composition.  Nonetheless, from UHECR data we know that SHDM can only make up for a small fraction of the UHECR flux below $E\approx10^{20}$~eV, but could come to dominate at higher energies (see, e.g., \cite{Marzola:2016hyt}).  UHECR data constrain the lifetime of SHDM to be larger than about $10^{22}$~yr to $10^{23}$~yr~\cite{Alcantara:2019sco}.  In this work we aim to show how ${\rm PBH \rightarrow SHDM}$ can in fact be tested with GWs in a distinctive manner.

Our idea is based on the following arguments: if, like all known elementary particles, the SHDM  acquires a mass via a phase transition (PT) mechanism,  the scale  of the transition (typically defined at a temperature $T\sim v$, where $v$ is the vacuum expectation value of the scalar field responsible for the PT) should be at least of the order of the SHDM mass.  The same PT, if it is strongly first-order~\cite{Mazumdar:2018dfl} or if it produces topological defects~\cite{cs1}, can also be a source of GWs.  It is well known that cosmic strings~\cite{cs1,cs2,cs3} which originate as topological defects due to the breaking of gauge symmetries such as $U(1)$, produce detectable GWs if the symmetry breaking scale $T\sim v\sim \Lambda_{CS}\gtrsim 10^{10}$~GeV~\cite{cs4,cs5,cs6,cs7,cs8,cs9,cs10,cs11}. Thus, cosmic strings provide an outstanding opportunity to probe the physics involving super-heavy particles~\cite{lepcs1,lepcs2,lepcs3,romepbh} that acquire a mass via PTs, because detectable GWs from cosmic strings can be naturally associated with the SHDM mass generation via the $U(1)$ breaking at a scale $\Lambda_{CS}\gtrsim M_{DM}$\footnote{From a more fundamental theory perspective, it is quite easy to naturally obtain a $U(1)$-symmetric SHDM mass term at the intermediate stage of a Grand Unified Theory (GUT) to SM  breaking, wherein the $U(1)$ can be identified as $U(1)_{B-L}$~\cite{lepcs1,bimal1,bimal2}.}.


The most distinct feature of GWs from cosmic strings is a scale invariant spectrum  spanning a wide range of frequencies, which makes them an ideal candidate for multi-frequency GW studies~\cite{cs9,cs10,cs11}.  If ultralight PBHs with large enough initial energy density existed in the early universe, apart from efficiently producing SHDM and diluting any preexisting SHDM density via entropy injection, they would introduce one further signature in the GW spectrum, namely a spectral break at a  frequency $f_*$, beyond which the spectrum turns red. The  frequency $f_*$  for a given SHDM mass is unique {\it provided} that the SHDM generated by the evaporation of PBHs makes up all of the observed DM. We find that, for the allowed range of SHDM masses, $f_*$ lies in a very constrained and testable range.  This peculiar feature in the GWs spectrum makes the ${\rm PBH \rightarrow SHDM}$ scenario quite distinct from standard  scenarios of GWs from cosmic strings\footnote{GWs can also arise directly from either PBH production mechanisms or their evolution~\cite{bhgw4a,bhgw5,bhgw6,Domenech:2021wkk}. Specifically, in Sec.\ref{sec5}, we discuss the induced GWs~\cite{bhgw6,Domenech:2021wkk} from the density fluctuations of PBHs which  potentially makes our scenario  unique in terms of its GW signatures.}.

Note also that in this set-up heavier SHDM  requires larger  symmetry breaking scales $\Lambda_{CS}$. Since the amplitude of GWs from cosmic strings increases with $\Lambda_{CS}$, the detectability improves with increasing SHDM mass. Adopting this point of view, it is tempting to interpret the recent finding of a  high-amplitude stochastic common-spectrum process across 45 pulsars by the NANOGrav pulsar-timing array~\cite{NANOGrav} (confirmed also by PPTA in an updated analysis~\cite{ppta}) in terms of a background of GWs radiated by CSs~\cite{nanofit1,nanofit2,nanofit3}.  In this case, we show that  the data can be fitted with a SHDM mass $3\times 10^{13}~\text{GeV} \lesssim M_{DM} \lesssim 10^{14}~\text{GeV}$, and that our model can be distinguished from other cosmic strings fits owing to its unique prediction of a spectral break at a turning-point frequency $0.2~\text{Hz} \lesssim f_* \lesssim 0.4~\text{Hz}$.

\begin{figure}
\includegraphics[scale=.6]{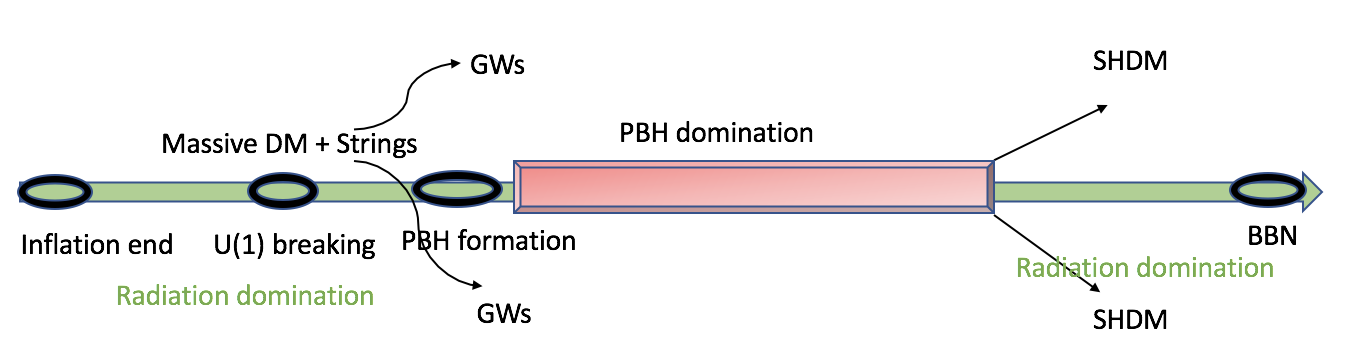}
\caption{A possible timeline for the proposed scenario.}\label{fig0}
\end{figure}

Let us briefly summarise the main ideas here.  We assume that ultra-light PBHs existed sometime in the very early Universe, before BBN.  These PBHs have a large initial energy density and eventually come to dominate the energy budget of the Universe before evaporating.  As they evaporate into all fields, they must also produce DM, which, as it turns out, has to be super-heavy.  If, and this is our second assumption, SHDM gets its mass via $U(1)$ symmetry-breaking, the phase transition that gives a mass to the DM also produces cosmics strings which radiate GWs.  The symmetry-breaking scale $\Lambda_{CS}$ has to be very large because SHDM is super-heavy.  Therefore, the GWs radiated by cosmic strings have a large enough amplitude to be detectable.  Lastly, the PBH domination epoch introduces a unique spectral break in the GW spectrum whose frequency is related to the SHDM mass.  Therefore, these marked features of a primordial GWs signal, potentially detectable across all frequencies accessible with current and planned experimental facilities, would indicate that SHDM originated from PBHs. A possible timeline for the proposed scenario is shown in Fig.\ref{fig0}.

The rest of the paper is organised as follows: in Sec.\ref{sec2} we obtain an expression for the GW background we expect from the cosmic string network.  In Sec.\ref{sec3} we explain how PBHs produce SHDM.  In Sec.\ref{sec4} we present our numerical results including the fit to NANOGrav data.  In Sec.~\ref{sec5} we discuss how GWs from PBH density perturbation could be a complementary signature of our model. We also discuss a possible extension of our scenario to address baryogenesis.  Finally, in Sec.\ref{sec6}, we conclude and give an outlook for further developments of this mechanism.

\section{Gravitational waves from cosmic strings and their spectral features}\label{sec2}

We consider a $U(1)$-charged complex scalar field $\Phi$ with the tree level potential $V(\Phi)=-\frac{m_\Phi^2}{2}|\Phi|^2+\frac{\lambda}{4}|\Phi|^4$, where $m_\Phi$ and $\lambda$ are the field's mass and self-interacting coupling constant, respectively.  After the spontaneous breaking of the $U(1)$ symmetry, the field $\Phi$  gets a vacuum expectation value ($\Lambda_{CS}\equiv v_\Phi= m_\Phi/\sqrt{\lambda}$) and generates the mass of the DM as well as gives rise to cosmic strings.  As a prototype, let the DM be a Majorana fermion\footnote{This, although not necessary for our purposes, is nonetheless useful in order to embed the scenario in a more realistic model which also includes baryogenesis, as we discuss in Sec.\ref{sec5}.} $\chi$ having $U(1)$ charge $q_\chi=-1$; therefore, to make the mass term $\mathcal{L}_{DM}\sim y_\chi\chi\chi\Phi$  symmetric under $U(1)$,  we set  $q_\Phi=2$.  Once the symmetry is broken the DM becomes massive ($M_{DM}=y_\chi v_\Phi$) and cosmic strings are formed. After their formation, the strings, which are randomly distributed in space, form close loops and a network of horizon-size long strings~\cite{ls1,ls2}. When two segments of long strings cross each other they inter-commute and form loops.  Long strings are characterised by a correlation length $L=\sqrt{\mu/\rho_\infty}$, where $\rho_\infty$ is the long string energy density and $\mu $ is the string tension defined as $\mu=\pi v_\Phi^2 h\left( \lambda/2 g^2\right)$. The quantity $h$ is a slowly varying function of its arguments ($\lambda$ and the gauge coupling $g$) with $h(1)\simeq 1$. In numerical simulations of cosmic strings, usually both the couplings are taken to be $\mathcal{O}(1)$ and therefore, the string tension $\mu$ is defined as $\mu=\pi v_\Phi^2$. We shall work with $y_\chi=\lambda=1$ and $g=1/\sqrt{2}$ so that the string tension becomes $\mu=\pi M_{DM}^2$. The choice $M_{DM}=v_\Phi$ is well motivated to make the model free from DM relics that may originate due to the pair production of SHDM from the scalar  ($M_\Phi=\sqrt{2}v_\Phi$) and the heavy gauge boson ($M_{Z}=v_\Phi$). In any case, at the end we will see that {\it any pre-existing SHDM relic} before PBH evaporation will be diluted significantly.

A string network interacts strongly with the thermal plasma, thereby  its motion gets damped~\cite{fric}. Once the damping phase is over, the strings oscillate and enter a phase of scaling evolution in which two competing dynamics coexist, namely, the stretching of the correlation length owing to the cosmic expansion and the fragmentation of the long strings into close loops which oscillate independently and produce GWs or particle radiation~\cite{cs4,cs5,cs6}. Between these two competing dynamics, there exists an attractor solution called the scaling regime~\cite{scl1,scl2,scl3} in which the characteristic length $L$ scales as cosmic time $t$. In this regime, for a constant string tension, we have $\rho_\infty\propto t^{-2}$. Therefore, the network tracks any cosmological background energy density $\rho_{bg}\propto t^{-2}$ with a small constant number proportional to $G\mu$, where $G$ is the Newton constant. This scaling behaviour prevents the cosmic string network from dominating the energy density of the Universe unlike all other cosmological defects.

Motivated by the results of cosmic string network simulations (see Refs.~\cite{cs7,cs8,scl1,scl2}) we always assume that the network is in the scaling regime while contributing to the GWs. The loops radiate GWs at a constant rate $\Gamma$ which determines the time evolution of a loop of initial (i.e., at creation time $t_i$) size $l_i=\alpha t_i$ as $l(t)=\alpha t_i-\Gamma G\mu(t-t_i)$, where $\Gamma\simeq 50$~\cite{cs4,cs6} and $\alpha\simeq 0.1$ are determined by numerical simulations~\cite{cs9,cs10}. The total energy loss from a loop can be decomposed into a set of normal-mode oscillations with frequencies $f_k=2k/l=a(t_0)/a(t)f$, where $k=1,2,3...k_{max}$ ($k_{max}\rightarrow\infty$), $f$ is the frequency observed at present time $t_0$ and $a$ is the scale factor of the Universe. The GW density parameter is given by $\Omega_{GW}(t_0,f)\equiv f\rho_c^{-1}d\rho_{GW}/df=\sum_k\Omega_{GW}^{(k)}(t_0,f)$, with the $k$-th mode amplitude $\Omega_{GW}^{(k)}(t_0,f)$ as~\cite{cs9} 
\bea
\Omega_{GW}^{(k)}(f)=\frac{2kG\mu^2 \Gamma_k}{f\rho_c}\int_{t_{osc}}^{t_0} \left[\frac{a(t)}{a(t_0)}\right]^5 n\left(t,l_k\right)dt,\label{gwf1}
\eea
where $n\left(t,l_k\right)$ is the loop number density, $\rho_c$ is the critical energy density of the Universe, $\rho_{GW}$ is the GW energy density and $l_k$ is loop length corresponding to the frequency $f_k$. The small scale structures of the loop sets the quantity $\Gamma_k$ which is given by $\Gamma_k=\frac{\Gamma k^{-\delta}}{\zeta(\delta)}$, with $\delta=4/3$ and $5/3$ for loops containing cusps and kinks, respectively~\cite{cuki}.  Eq.\ref{gwf1} is valid only for $t_i>t_{osc} = {\rm Max}\,\left[{\rm network~formation~time}~(t_F),{\rm end~of~damping}~(t_\text{fric})\right]$ and $t_i > l_{crit}/\alpha$, with $l_{crit}$ the critical length above which GWs dominate over the massive particle radiation~\cite{partrad1,partrad2}. As discussed in Sec.\ref{sec5}, both these bounds set a high-frequency cut-off in the spectrum (a comprehensive analysis can be found in Ref.~\cite{spb2}). 

With the standard cosmological evolution,  Eq.\ref{gwf1} leads to the following spectral features: (I) a peak at low frequency due to the GWs from the loops that are produced in the radiation era and decay in the following matter era and (II) a plateau at high frequency \cite{cs9,cs10,vos3} given by (see Appendix \ref{appa})
\bea
\Omega_{GW}^{plt}(f)=\frac{128\pi G\mu}{9\zeta(\delta)}\frac{A_r}{\epsilon_r}\Omega_r\left[(1+\epsilon_r)^{3/2}-1\right], \label{flp1}
\eea
where $\epsilon_r=\alpha/\Gamma G\mu \gg 1$, $A_r\simeq 5.4$ being the loop production efficiency in radiation domination and $\Omega_r\sim 9\times 10^{-5}$ is the present radiation energy density fraction---this behaviour arises from the loop dynamics in the radiation era. Since $\mu\sim \Lambda_{CS}^2$, the plateau amplitude is $\Omega_{GW}^{plt}\sim \Lambda_{CS}$: this is a property that makes cosmic strings an outstanding probe of super-high-scale physics---in our context this scale is related to the SHDM mass scale.

For a very long period of radiation domination, the plateau remains intact across all frequencies accessible by current and planned GW detectors such as LISA~\cite{LISA} and LIGO~\cite{ligo1,ligo2}. However, if a non-standard cosmological evolution is introduced before the onset (at $T_*$) of the radiation domination era that leads to BBN ($T_\text{BBN}\sim 5$ MeV)~\cite{bbn,bbn1,bbn2}, the spectrum deviates from the constant plateau at high frequencies: {\it this happens at a turning-point frequency $f_*$ which depends on the end of the non-standard evolution epoch}. In case of an early matter domination, the spectral slope changes from $\Omega_{GW}\sim f^0$ to $\Omega_{GW}\sim f^{-1}$ beyond $f_*$. The sensitivity reach of GW detectors to probe $f_*$ and $T_*$ can be found, e.g., in Fig.~4 and~5 of Ref.~\cite{cs11}. This scenario is relevant in our case, since the PBHs introduce an early matter domination era before BBN. For the $k=1$ mode, the $f_*$ can be calculated in a simple analytical way that gives~\cite{lepcs3,cs11}
\bea
f_*=\sqrt{\frac{8}{\alpha\Gamma G\mu}}t_*^{-1/2}t_0^{-2/3}t_{\rm eq}^{1/6}\simeq \sqrt{\frac{8 z_{\rm eq}}{\alpha\Gamma G\mu}}\left(\frac{t_{\rm eq}}{t_*}\right)^{1/2}t_0^{-1},\label{br0}
\eea
where $z_{\rm eq}\simeq 3387$ is the red-shift at the usual matter-radiation equality, which takes place at time $t_{\rm eq}$. \\

Note that even though the GW spectrum beyond the turning-point frequency behaves as $\Omega(f>f_*)\sim f^{-1}$ for $k=1$, when the contributions from all the modes are taken into account one finds $\Omega(f>f_*)\sim f^{-1/3}$ (see Appendix \ref{appb}). Moreover, even without a non-standard cosmological epoch, GWs from cosmic strings may exhibit a non-flat spectrum in the case of global strings~\cite{Gorghetto:2021fsn,globst}, inflation-diluted strings~\cite{csinf1,csinf2} and melting strings~\cite{meltst}.  

\section{ generalities of PBH dynamics and Dark matter  production}\label{sec3}

\begin{figure}
\includegraphics[scale=.35]{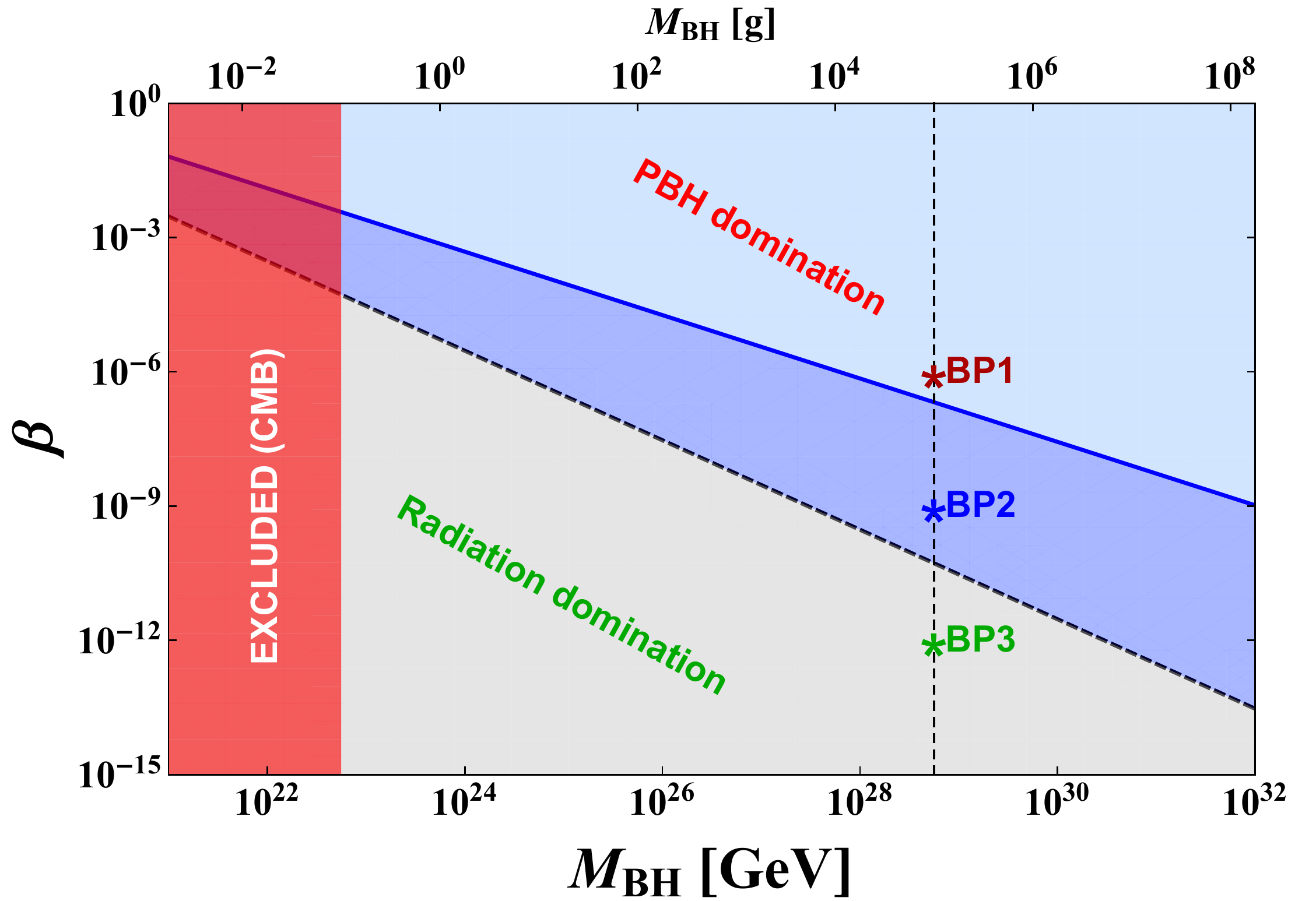}\includegraphics[scale=.33]{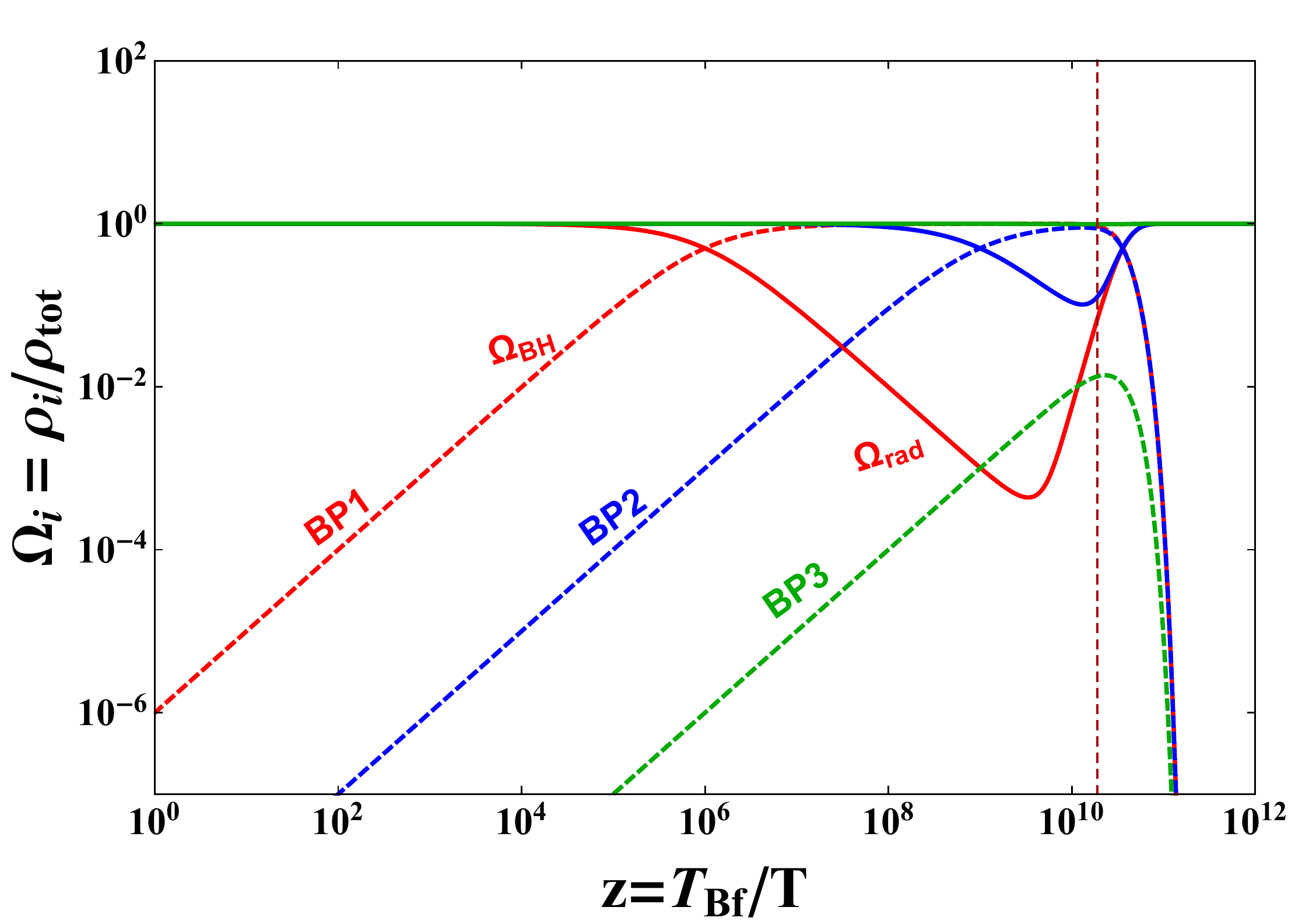} 
\caption{Left panel: the parameter $\beta$ of Eq.\ref{beta_def} as a function of $M_{BH}$. The region above (below) the dashed black line corresponds to PBH (radiation) domination. The region between the dashed black and solid blue lines represents the  allowed ranges of $\beta$ (cf. Ref.\cite{bhgw6}). The horizontal red  region is excluded by the CMB constraint on Hubble scale of inflation $H_{\rm inf}\lesssim 3\times 10^{14}$~GeV at 95$\%$ CL~\cite{Planck}.  Right panel: evolution of the energy density fractions of the PBHs (dashed lines) and radiation (solid lines) for $M_{BH}=10^5$~g. In both plots the benchmark points correspond to $\beta=10^{-6,-9,-12}$. The dashed vertical line corresponds to the PBH evaporation temperature calculated analytically in Eq.\ref{teva} for radiation domination.  Eq.\ref{tevab} would correspond to a very similar line.} \label{fig1}
\end{figure}

A useful parameter to track the evolution of the  energy density of the black holes is 
\begin{align}\label{beta_def}
    \beta\equiv \frac{\rho_{ BH}(t_{Bf})}{\rho_{\rm R}(t_{Bf})} \,,
\end{align}
where $\rho_{\rm R}(t_{Bf})$ and $\rho_{BH}(t_{Bf})$ are the energy densities of PBHs and radiation at the PBH formation time $t_{Bf}$, respectively. 
If the PBHs do not dominate the energy density of the Universe at their evaporation, the parameter $\beta$ is bounded from above as~\cite{bhgw4a} (Appendix \ref{appc}) 
\bea
\beta<\gamma^{-1/2}\left(\frac{\mathcal{G }g_{*B}(T_{BH})}{10240\pi}\right)^{1/2}\frac{M_{Pl}}{M_{BH}},\label{beta}
\eea
where $\gamma\simeq 0.2$ is the black hole formation efficiency, $T_{BH}=M_{Pl}^2/8\pi M_{BH}$~\cite{hr} is the PBH temperature, $\mathcal{G}\simeq 3.8$ being the graybody factor, $g_{*B} \simeq 100$ is the number of relativistic particle species below $T_{BH}$ in the SM with three left-handed light neutrinos~\cite{MacGibbon:1991tj} and we approximate non-rotating PBHs with a monochromatic mass spectrum. In Fig.~\ref{fig1} (left panel) we show the $\beta$ parameter as a function of the PBH mass. The region above (below) the dashed black line corresponds to PBH (radiation) domination. The region between the dashed black and solid blue lines is allowed for the induced GWs by PBH density fluctuation not to saturate the BBN bound\cite{bhgw6} (see discussion in Sec.\ref{sec5}). For future reference, we also indicate three benchmark values BP$(1,2,3)=10^{-6,-9,-12}$ for $M_{BH}=10^5$~g. Note that  BP3 is in the radiation domination region, i.e., for $\beta=10^{-12}$, PBHs of mass $10^{5}$~g can never dominate the energy density of the Universe. On the contrary, PBHs come to dominate for some time for the other two BPs.\\

Larger values of $\beta$ correspond to longer period of PBH domination: As we discuss later in this section, a large value of $\beta$ is preferred in our model  to make ${\rm PBH\rightarrow SHDM}$ the dominant channel. The time evolution of the energy densities can be understood by solving the following Friedmann equations\footnote{These two equations are derived from the standard Friedmann equations in presence of PBHs and the entropy non-conservation equation due the evaporation of PBHs, see Appendix \ref{appc}. }~\cite{dm5a,Giudice:2000ex}

\bea
\frac{d\rho_{R}}{dz}+\frac{4}{z}\rho_R&=&0\label{den1}\\
\frac{d\rho_{BH}}{dz}+\frac{3}{z}\frac{H}{\mathbb{K}}\rho_{BH}&-&\frac{\dot{M}_{BH}}{M_{BH}}\frac{1}{z\mathbb{K}}\rho_{BH}=0,\label{den2}
\eea
where $z=M_0/T$ with $M_0$ being an arbitrary mass scale which we fix as $M_0 = T_{Bf}$ (PBH formation temperature) and the parameter $\mathbb{K}$ defined as
\bea
\frac{1}{T}\frac{dT}{dt}=-\left(H+\frac{1}{3g_{*s}(T)}\frac{dg_{*s}(T)}{dt}+\frac{\dot{M}_{BH}}{M_{BH}}\frac{\rho_{BH}}{4\rho_{R}}\right)\equiv-\mathbb{K}\label{temvar}
\eea
with $ g_{*s}(T)$ being the number of entropy degrees of freedom. The solutions of Eq.\ref{den1} and Eq.\ref{den2} are shown in the right panel of Fig.~\ref{fig1} for the same BPs as in the left panel. It can be seen that, the larger $\beta$, the longer the period of PBH domination. Even though the non-thermal DM production from PBHs is independent of $\beta$ provided PBHs dominate the energy density of the Universe, large values of $\beta$ make this scenario more predictive as any preexisting relic DM density will be mostly washed out, making the ${\rm PBHs\rightarrow DM}$ channel the dominant one.\\

Let us now turn to the production of DM from PBHs. A general expression for the DM relic energy density fraction is given by
 \bea
\hspace{-0.7cm}
\Omega_{\rm DM}h^2=\frac{M_{\rm DM} n_\gamma^0}{10.54 f({ T_{\rm ev}}, T_0){\rm GeVm^{-3}}} \left(\frac{N_{{\rm DM}}}{N_\gamma}\right)_{ T_{\rm ev}} \simeq 1.45 \times 10^6 \left(\frac{N_{{\rm DM}}}{N_\gamma}\right)_{ T_{\rm ev}} \left( \frac{M_{\rm DM}}{\rm GeV}\right),\label{relic}
\eea
where $N_{\rm DM}$ and $N_{\gamma}$ are the DM and photon number densities, respectively, normalised to the ultra-relativistic equilibrium number density $n_{f,\rm eq}^{\rm ur}=g_{f}T^3/\pi^2$ of a spin-$1/2$ fermion\footnote{This is a standard `$T^3$' normalisation. One can also normalise it with respect to the entropy density $s\sim T^3$~\cite{dm5a}.}, $n_\gamma^0\simeq 410.7\times 10^{6} \rm m^{-3}$ and $f( T_{\rm ev}, T_0)\simeq 27.3$ are the relic photon number density at the present time and photon dilution factor respectively. Given that $\Omega_{\rm DM}h^2\simeq 0.12$~\cite{Planck:2015fie}, from Eq.\ref{relic} we find
\bea
N_{\rm DM}^{\rm ev}= N_{\rm DM}^{\rm Obs}\simeq 1.1\times 10^{-7} \left(\frac{\rm GeV}{ M_{\rm DM}} \right). \label{NDM}
\eea
One has to compare Eq.\ref{NDM} with the number density of DM produced by PBHs, i.e.,  
\bea
N_{\rm DM}^{\rm ev}=N_{\rm BH}^{\rm ev}\bar{n}_{\rm DM},\label{DMobs}
\eea
where $\bar{n}_{\rm DM}$ is the number of DM particles produced by a black hole and $N_{\rm BH}^{\rm ev}$ is the normalised number density of PBHs at evaporation. \\

Computing $N_{\rm BH}^{\rm ev}\bar{n}_{\rm DM}$ in a PBH dominated Universe (see Appendix \ref{appd}) and noting that $N_{DM}^{\rm ev}\lesssim N_{DM}^{Obs}$ we obtain
\bea
M_{DM}\lesssim 3\times 10^{-7}\left(\frac{M_{BH}}{M_{Pl}}\right)^{1/2}{\rm GeV}~~{\rm for}~~T_{BH}>M_{DM},\label{ldm}
\eea
and 
\bea
M_{DM}\gtrsim 4.5\times 10^3\left(\frac{M_{BH}}{M_{Pl}}\right)^{-5/2}M_{Pl}^2~~{\rm GeV^{-1}}~~{\rm for}~~T_{BH}<M_{DM}\label{hdm}
\eea
where the sign $\simeq$ corresponds to $\Omega_{\rm DM}h^2\simeq0.12$ and in Eq.\ref{hdm}, dark matter is produced when the PBHs becomes light enough so that the corresponding $T_{BH}$ becomes larger than  $M_{DM}$ (see Appendix \ref{appd}).  Eq.\ref{ldm} and Eq.\ref{hdm} represent to the light and the heavy DM scenarios, respectively.  However, there is another constraint that needs to be taken into account in the case of non-thermal light DM. The constraint comes from the fact that the free-streaming length of light DM should not be too large at $z_{\rm eq}$ (Ly$\alpha$ cloud constraint) or it will interfere with structure formation~\cite{lmfs1,lmfs2,lmfs3,lmfs4}. This has been studied extensively in~\cite{dm2,dm5a,dm6}, and the result is that the DM mass is constrained from below as
\bea
M_{DM}>4.4\times10^{-6}\left(\frac{M_{BH}}{M_{pl}}\right)^{1/2}.\label{frst}
\eea

\begin{figure}
\includegraphics[scale=.45]{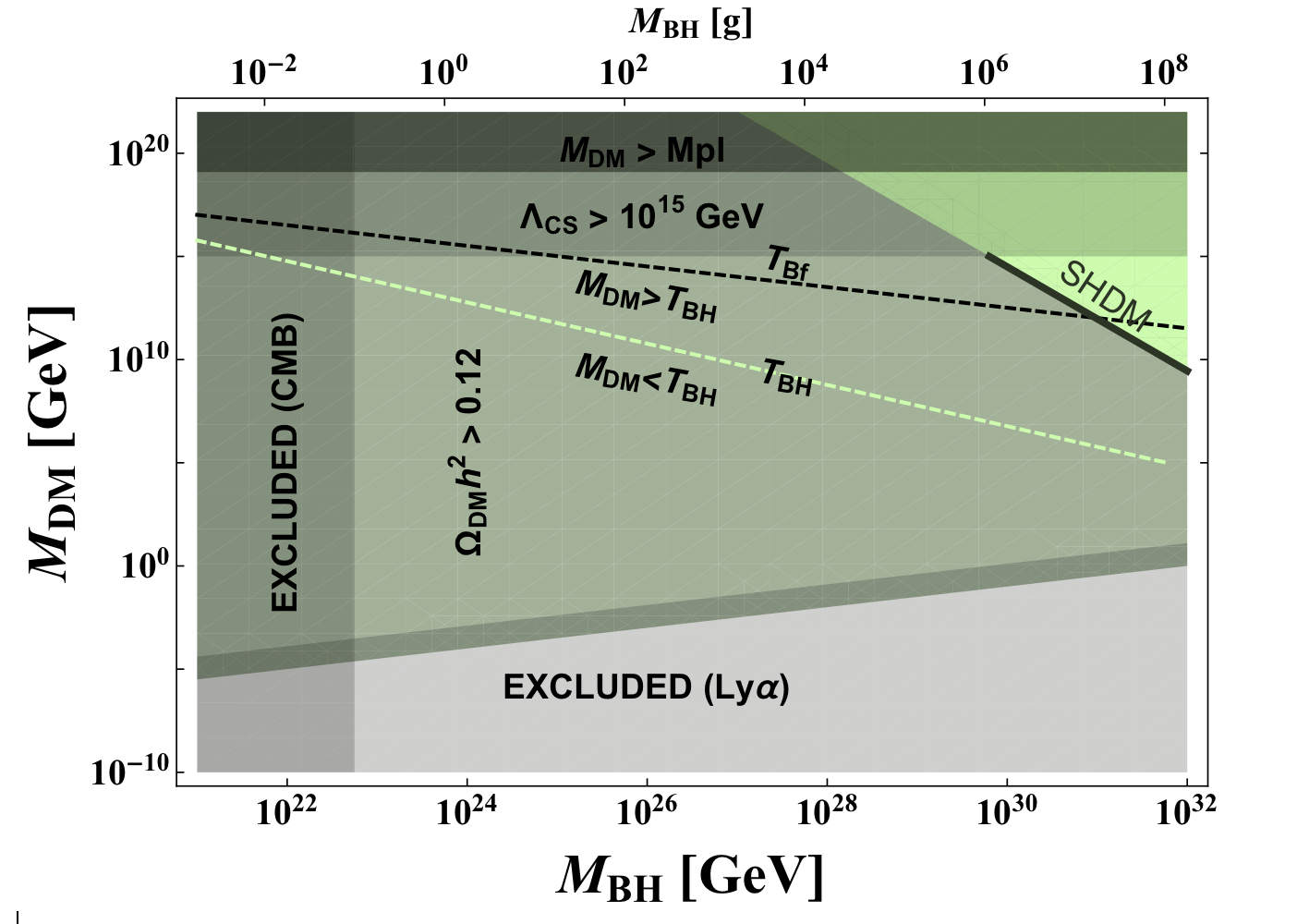} 
\caption{Parameter space in the $M_{DM}-M_{BH}$ plane: the light green region in the top-right corner is the allowed region ($\Omega_{DM}h^2<0.12$) and the thick solid line marks the values for which we obtain the observed dark matter abundance $\Omega_{DM}h^2\simeq 0.12$. We exclude $M_{DM}>10^{15}$~GeV assuming that the symmetry breaking that generates the dark matter mass takes place after inflation, which also implies that the cosmic string loop number density will be undiluted. The gray region is excluded due to the too large free-streaming length of the DM.}\label{fig2}
\end{figure}

\begin{figure}
\includegraphics[scale=.45]{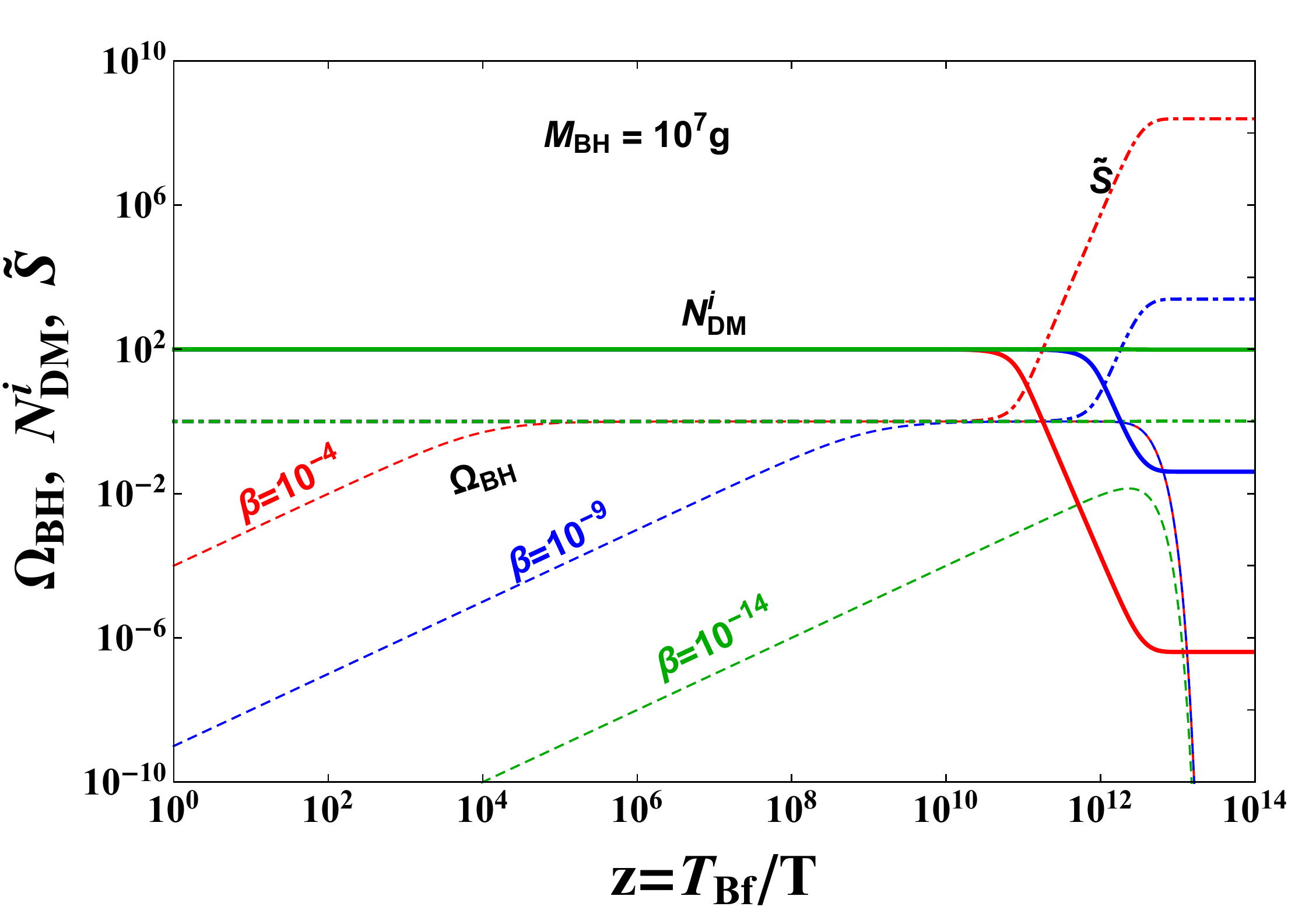} 
\caption{The evolution of the PBH energy density fraction (dashed lines), a preexisting dark matter abundance (solid lines) and the total entropy density (dot-dashed lines) for a long PBH-domination (red) or short PBH-domination (blue) period, or in radiation domination throughout (green).}\label{fig3}
\end{figure}

The constraint in Eq.\ref{frst} is incompatible with that in Eq.\ref{ldm} and rules out the ${\rm PBHs\rightarrow DM}$ scenario for light DM. Therefore the only viable option is Eq.\ref{hdm}, i.e., the heavy (in fact, super-heavy) DM scenario. All the constraints on the $M_{DM}-M_{BH}$ parameter space are shown in Fig.~\ref{fig2}. In general, the light green region in the top-right corner is allowed, i.e., for this region one has $\Omega_{DM}h^2<0.12$. Note that, the produced DM is super-heavy and in principle the mass $M_{DM}$ spans the range  $10^{10}{\rm GeV}\lesssim M_{DM}\lesssim M_{Pl}$. This range is much wider than what can be achieved in scenarios of gravitational production~\cite{shdm1,shdm2,shdm3}. However,  conservatively, we will limit ourselves to a maximum mass of $M_{DM}^{max}\simeq 10^{15}\sim T_{RH}^{max}$~GeV, i.e., a post-inflationary symmetry breaking that generates the DM mass as well as an undiluted cosmic strings loop number density~\cite{csinf1,csinf2}. The upper limit on $M_{DM}$ corresponds to a lower bound on the PBH mass: $M_{BH}\gtrsim 10^6$~g. \\

Let us conclude this section by analysing the fate of a preexisting DM relic ($N_{DM}^i$) for the three benchmark values of $\beta$ and for $M_{BH}=10^7$~g, which corresponds to $M_{DM}\simeq 5\times 10^{12}$~GeV (cf.\ Fig.~\ref{fig2}). 
 First of all, in presence of PBHs, the Boltzmann equation for a preexisting DM relic can be derived as
\bea
\frac{dN_{DM}}{dz}-\frac{3}{\mathbb{K}z}\left(\mathbb{K}-H\right)N_{DM}=0\label{pre}
\eea
which in the case of $\mathbb{K} \rightarrow H$, i.e., when the PBHs are absent or do not dominate the energy density of the universe, leads to $N_{DM} (z\rightarrow\infty)=N_{DM}^{i}$, as expected.  However, when PBHs dominate for some period the situation is very different: due the appearance of the second term in the LHS of Eq.\ref{pre}, any preexisting relic will be diluted by an amount that depends on $\beta$ and hence on the duration of the PBH domination epoch. The final value of $N_{DM}^{i}$ is then given by
\bea
N_{DM}^f=N_{DM}^i{\rm exp}\left[ \int_1^\infty \frac{3}{\mathbb{K}z}\frac{\dot{M}_{BH}}{M_{BH}}\frac{\rho_{BH}}{4\rho_R}\right].
\eea
This dilution follows from the entropy non-conservation equation (Eq.\ref{be3}) that leads to $\mathbb{K}\neq H$ in Eq.\ref{temvar}. Neglecting the time variation of $g_{*s}$, the amount of total entropy $\tilde{S}\sim a^3/z^3$ production can be tracked using 
\bea
\frac{da}{dz}=\left(1-\frac{\dot{M}_{BH}}{M_{BH}}\frac{\rho_{BH}}{4\rho_{R}\mathbb{K}}\right)\frac{a}{z}
\eea
that can be easily derived from Eq.\ref{be3}.

In Fig.~\ref{fig3}, we show the evolution of $N_{DM}^i(z)$, $\Omega_{BH}(z)$ and $\tilde{S}(z)$ in solid, dashed and dot-dashed lines respectively; we show the cases $\beta=10^{-4}$ (red, strong PBH domination), $\beta=10^{-9}$ (blue, mild PBH domination), $\beta=10^{-14}$ (green, radiation domination).  Note that the longer the PBH domination, the more the produced entropy, and the stronger is the dilution of $N_{DM}^i$. For example, a PBH of mass $10^7$~g with $\beta=10^{-4}$ can dilute any preexisting DM density by a factor of $10^8$. However, the amount of a preexisting relic is strongly model dependent, e.g., in our case, if kinematically allowed, the $U(1)$~gauge boson or scalar can produce the initial $N_{DM}^i$. Nonetheless, as shown already, in case of a strong PBH domination such a relic density is strongly suppressed at the time of PBH evaporation. In scenarios like those of Refs.~\cite{bhgw5,bhgw6}, where the GWs that come from the density perturbations of PBHs can potentially give an upper bound on $\beta$, a large value of $N_{DM}^i$ may be a cause of concern.

\section{Imprints of SHDM on the GWs from  DM mass origin/cosmic strings }\label{sec4}

\begin{figure}
\includegraphics[scale=.37]{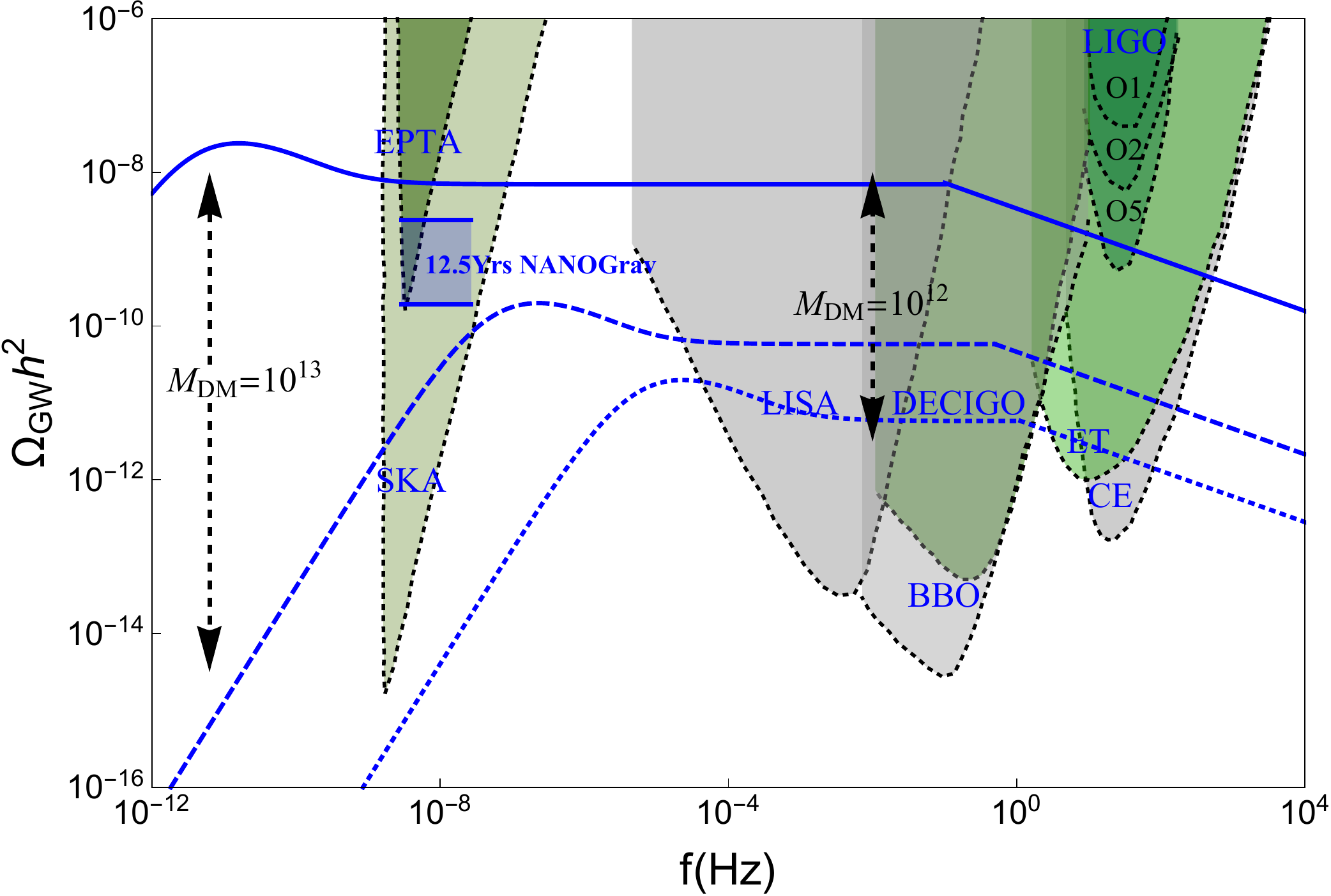}\includegraphics[scale=.37]{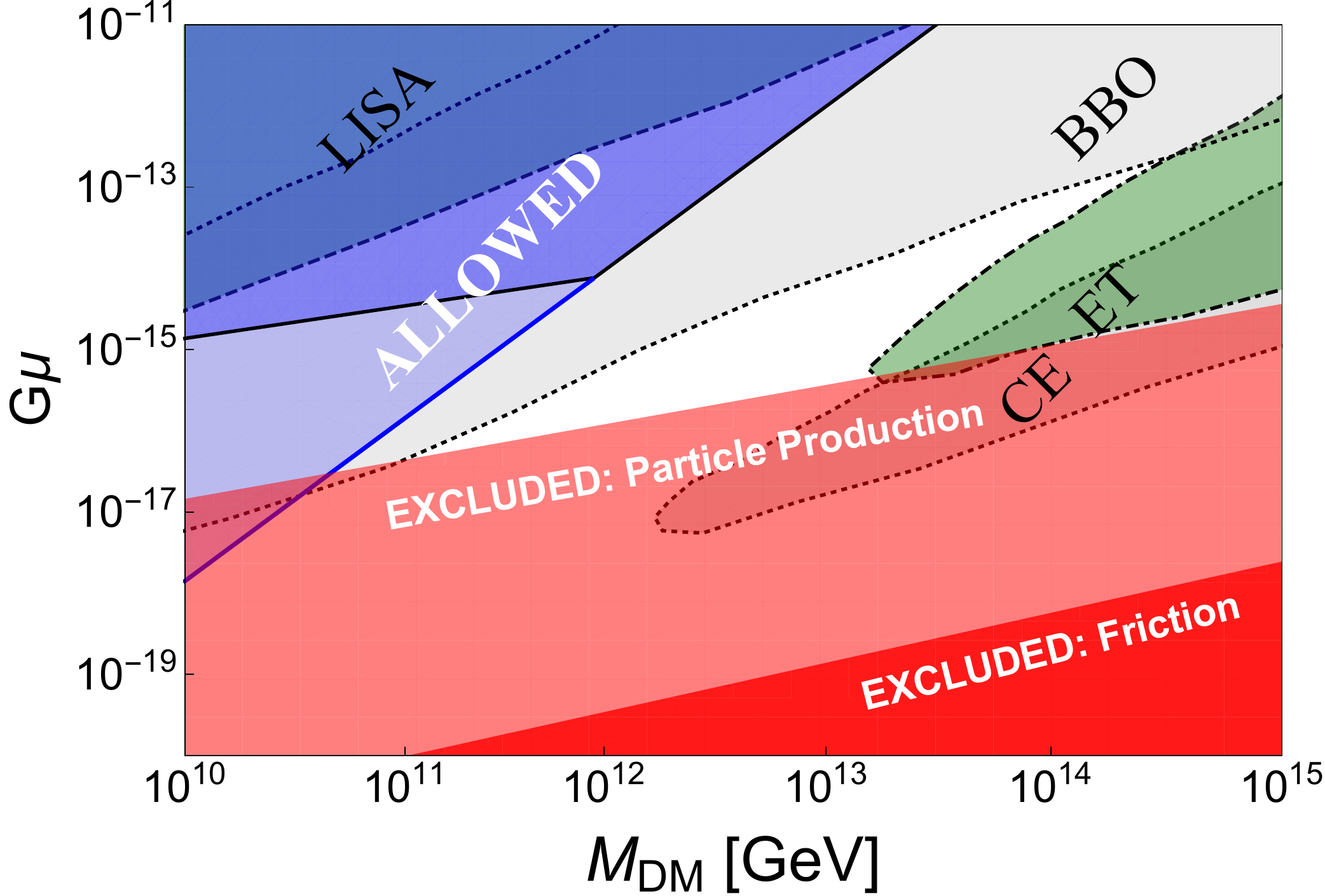} 
\caption{Left panel: the gravitational wave spectrum for our $\rm{PBHs\rightarrow SHDM }$ scenario. The solid blue line represents the spectrum for $M_{DM}\simeq 10^{15}$~GeV. As the SHDM mass decreases the parameter space on the $\Omega_{GW}h^2-f$ plane enlarges, the allowed region for $M_{DM}\simeq 10^{15}{\rm GeV}-10^{13}$~GeV lies between the solid and the dashed lines, whereas the region between the solid and the dotted lines is allowed for $M_{DM}\simeq 10^{15}{\rm GeV}-10^{12}$~GeV. Right panel: for a given value of $G\mu$ we show the potential of various detectors to probe the PBH evaporation temperature and hence $M_{DM}$. The blue regions are the allowed parameter space in our ${\rm PBHs\rightarrow SHDM} $ scenario. The red shades are excluded due  particle production from the loops and thermal friction that freezes the network, see main text.}\label{fig4}
\end{figure}

We start from the scenario in which the SHDM becomes massive before the PBHs are formed (Fig.\ref{fig0}).  The DM mass scale corresponding to correct DM density is larger than the PBH formation temperature for $M_{DM}\gtrsim M_{DM}^c\simeq 10^{12}$~GeV (Fig.~\ref{fig2}, thick black line). This leads to a symmetry breaking scale  $v_\Phi \equiv\Lambda_{CS} \simeq {\rm Max}\left[T_{Bf},M_{DM}\right]$. To evaluate $G\mu$, one can therefore set $\mu \sim M_{DM}^2$ for $M_{DM}>M_{DM}^c$ and $\mu \sim T_{Bf}^2$ for $M_{DM}<M_{DM}^c$\footnote{Note that $\mu \sim T_{Bf}^2$ is a possibility (Fig.\ref{fig0}), in which case, the condition $M_{DM}\lesssim T_{Bf}$ requires $y_\chi\lesssim 1$. Because we discuss only the $\mu \sim M_{DM}^2$ scenario, we shall not explore this possibility further.}. As mentioned earlier, in our set-up, the GW amplitude as well as the spectral features are sensitive to the details of the ${\rm PBHs\rightarrow SHDM}$ mechanism. Let us look at the amplitude first. Consider the case $v_\Phi=M_{DM}>10^{12}$ GeV.   The spectra of GWs for $M_{DM}=10^{15,13,12}$ GeV are shown in Fig.~\ref{fig4} (left panel) in solid, dashed, and dotted blue lines, respectively. The spectrum is in accordance with the standard expectation that with large $v_\Phi$ the amplitude of GWs increases, i.e., in our case, the prospects for detection improve as the SHDM becomes more massive. The most interesting fact is that even if the SHDM mass is as `light' as $10^{11}$~GeV, mid-band detectors such as LISA~\cite{LISA}, DECIGO~\cite{DECIGO} and BBO~\cite{BBO} are able to test the parameter space fully. \\

In principle we can relax the assumption of $y_\chi=1$, e.g., a SHDM of mass $M_{DM}\sim 10^{14}$~GeV can be obtained with a symmetry breaking scale $v_{\Phi}\sim 10^{15}$~GeV  when $y_\chi=0.1$. In this case, the string tension $G\mu$ ranges from $\sim Gv_\Phi^2$ to $\sim G M_{DM}^2$. Therefore, as we go to lower DM masses (compared to $v_\Phi$), the allowed range of $G\mu$ increases. We indicate this in Fig.~\ref{fig4} (left panel).  For $M_{DM}=10^{13}(10^{12})$~GeV the GW spectrum can be anywhere within the region between the solid and the dashed (dotted) lines. In the case of $y_\chi\ll 1$, the SHDM could be produced by the scalar and the gauge boson having abundant initial number densities. In that case, one has to consider  the period of PBH domination accordingly to dilute the generated $N_{DM}^i$. In this article, we do not present any explicit computation in this direction.\\

We turn now to the spectral shape by computing the turning-point frequency $f_*$. First of all, the PBH evaporation temperature in the case where they dominate the energy density, is derived as (see Appendix \ref{appd}):
\bea
T_{ev}=\left(\frac{5 M_{Pl}^2}{\pi^3 g_*(T_{ev})\tau^2}\right)^{1/4},\label{tevab1}
\eea
with $\tau$ being the  life-time of a PBH. Therefore, from  Eq.\ref{tevab1} and Eq.\ref{hdm} we obtain the  evaporation temperature in terms of the SHDM mass as
\bea
T_{ev}=2.1\times 10^{-8} \left(\frac{M_{DM}}{{\rm GeV}}\right)^{3/5}.\label{tevdm}
\eea
From Eq.\ref{br0}, the turning point frequency $f_*$ can then be expressed in terms of $T_{ev}$ and hence the SHDM mass as
\bea
f_* \simeq 2.1\times 10^{-8}\sqrt{\frac{50}{z_{\rm eq}\alpha \Gamma G\mu}}\left(\frac{M_{DM}}{T_0}\right)^{3/5}T_0^{-2/5}t_0^{-1},\label{dmbr}
\eea
where $T_0=2.7$ K is the photon temperature today.  {\it The expression~\ref{dmbr} is of immense interest and it is one of the most important results of this work}. For a given value of the DM mass and $G\mu$ (for $y_\chi=1$, $G\mu\sim GM_{DM}^2$), one can determine approximately at which frequency the spectrum changes slope from a plateau described by $f^0$ to $f^{-1/3}$ (see the discussion in Sec.\ref{sec2}). For example, a SHDM of mass $10^{15}$~GeV would correspond to a break at $f_*\sim 0.1 $~Hz, as shown in Fig.~\ref{fig4} (left panel). A robust prediction of our ${\rm PBHs\rightarrow SHDM}$ set-up is that, given the allowed range of $M_{DM}$ (Fig.~\ref{fig2}) , {\it the turning-point frequencies always lie within the sensitivity range of mid-band detectors:} $0.1~\text{Hz} \lesssim f_* \lesssim 7~\text{Hz}$. This window will enlarge for $y_\chi\ll 1$.  Reconstruction techniques for such  spectral shapes can be found in~\cite{Caprini:2019pxz}. \\

In Fig.~\ref{fig4} (right panel), we show the allowed range in the $G\mu-M_{DM}$ parameter space that corresponds to such spectral breaks. The dark-blue region corresponds to the case when $v_\Phi>T_{Bf}$ (see Fig.\ref{fig0}), whereas the parameter space enlarges to include the light-blue region  when the condition $v_\Phi>T_{Bf}$  is relaxed. As mentioned earlier in Sec.\ref{sec2}, the loops effectively contribute to the GWs for $t>t_i$ which is bounded from below in presence of thermal friction and particle production from the loops. A lower bound on $t_i$ (upper bound on temperature) also sets a cut-off on the GW spectrum at a high frequency beyond which the spectrum falls. The effects of thermal friction is sub-dominant when $H\sim T^2/M_{Pl}>\Gamma_{fric} \sim T^3 / \mu$~\cite{fric}. This implies that  the motion of the network gets damped until $T_{fric}\sim \mu/M_{Pl}$. For the  GW spectrum to fall due to PBH domination, which is  our requirement, one should have $T_{fric}>T_{ev}$. This corresponds to the constraint 
\bea
G\mu > 2.1\times 10^{-8} \left(\frac{M_{DM}}{{\rm GeV}}\right)^{3/5} \left(\frac{M_{Pl}}{\rm GeV}\right)^{-1}.\label{fb}
\eea 

A very recent numerical simulation reported that there is a critical size $l<l_{crit}$ for which particle production is the dominant radiation mode and the loop number density does not scale~\cite{partrad1,partrad2}.  This corresponds to a lower bound  $t_i>l_{crit}/\alpha$ and therefore a cut-off on the GW spectrum at a high frequency. Now proceeding in the same way as in the friction case, a bound on $G\mu$ is obtained as~\cite{lepcs3}
\bea
G\mu >1.7\times 10^{-22}\left(\frac{M_{DM}}{{\rm GeV}}\right)^{12/25},\label{pb}
\eea
where we have considered particle production from the cusps  with $l_{crit}=\frac{\mu^{-1/2}}{(\Gamma G \mu)^2}$~\cite{partrad1,partrad2} and $\alpha=0.1$. The constraints in Eq.\ref{fb} and Eq.\ref{pb} are shown in Fig.\ref{fig4} (right) by dark-red and light-red shades respectively. It is evident that the particle production cut-off is much stronger and rules out a portion of the allowed parameter space.\\

We now discuss how, in this set-up, the recent finding of a stochastic common spectrum process by the NANOGrav pulsar timing array would constrain the range of SHDM mass and how the turning point frequencies potentially can be a complementary probe of our ${\rm PBHs\rightarrow SHDM}$ scenario.  With their recently released 12.5 yrs data set, the NANOGrav collaboration has reported strong evidence for a stochastic common-spectrum process across 45 pulsars~\cite{NANOGrav}. Nonetheless, since the time residuals do not show the characteristic angular correlation described by the Hellings–Downs curve~\cite{Hellings:1983fr}, the detection has not been claimed as GWs. Moreover, systematics such as solar system effects~\cite{solar} and pulsar spin noise~\cite{spin} may affect the signal.\\

\begin{figure}
\includegraphics[scale=.45]{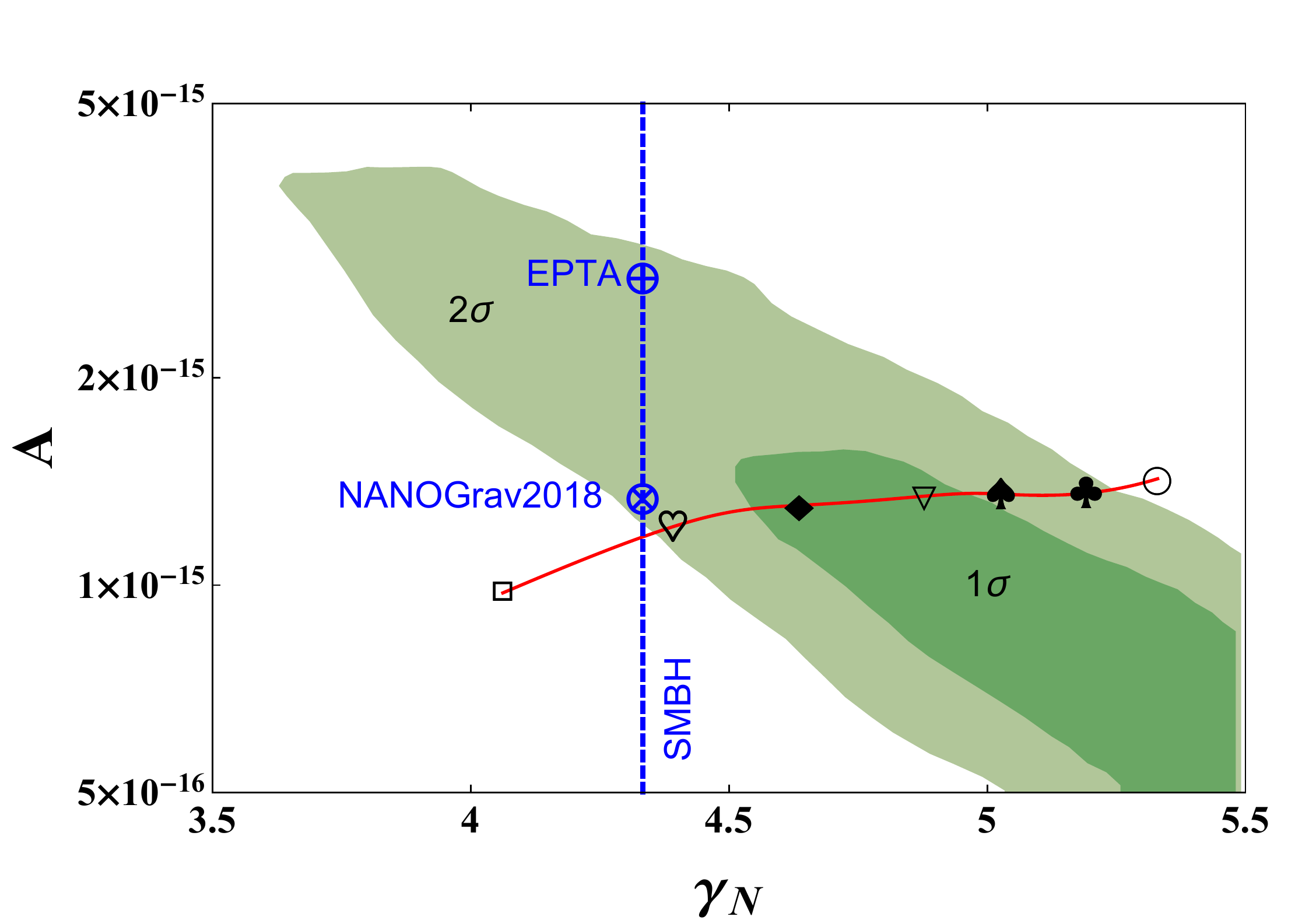} \includegraphics[scale=.25]{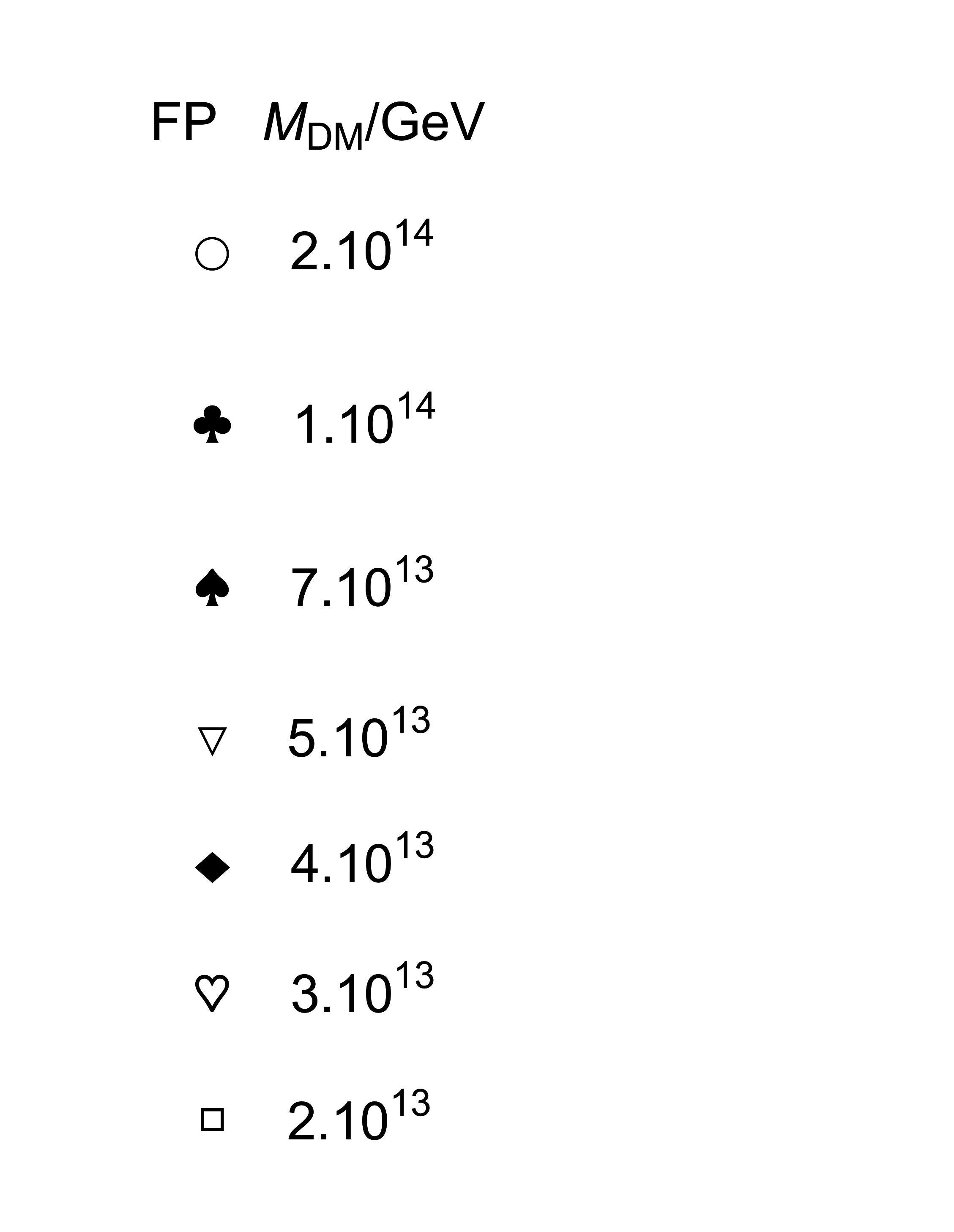}
\caption{A fit to the NANOGrav data for $2\times 10^{13}~{\rm GeV}\lesssim M_{DM} \lesssim 2\times 10^{14}~{\rm GeV}$ (red line). Points marked with different shapes correspond to the different values of the SHDM mass reported in the legenda to the right of the plot. The NANOGrav$@2\sigma$ range is well fitted with a SHDM mass within $3\times 10^{13}~{\rm GeV}\lesssim M_{DM} \lesssim 1 \times 10^{14}~{\rm GeV}$.}\label{fig5}
\end{figure}
Propitiously, if the signal is interpreted as GWs, cosmic strings provide an excellent explanation of such a finding~\cite{nanofit1,nanofit2,nanofit3}.  Let us mention that, while the NANOGrav 12.5~yr data~\cite{NANOGrav} is nearly consistent with previous EPTA data~\cite{Lentati:2015qwp} and with a very recent analysis of the PPTA data set~\cite{ppta}, the older NANOGrav 11~yrs data set~\cite{Arzoumanian:2018saf} is in tension with the new one. Ref.~\cite{NANOGrav} mentions that, with the older data, the tension would be reduced once an improved prior for the intrinsic pulsar red noise will be used. The NANOGrav common-spectrum process is expressed in terms of a power-law signal with characteristic strain given by 
\bea
h_c(f)=A\left(\frac{f}{f_{yr}}\right)^{(3-\gamma_N)/2}
\eea
with $f_{yr}=1~{\rm yr}^{-1}$, $A$ and $\gamma_N$ being the characteristic strain amplitude and the timing-residual cross-power spectral index ($\gamma_N=13/3$ for super-massive black hole mergers), respectively. The abundance of GWs can be recast in the standard form as:
\bea
\Omega_{GW}(f)=\frac{2\pi^2}{3H_0^2}f^2h_{c}(f)^2=\Omega_{yr}\left(\frac{f}{f_{yr}}\right)^{5-\gamma_N}, ~~~~~{\rm with}~~~~~ \Omega_{yr}=\frac{2\pi^2}{3H_0^2}A^2f^2_{yr}.\label{po}
\eea

The power law approximation was fitted to 5 bins covering approximately the frequency range $f\in \left[2.5\times 10^{-9}, 1.2\times10^{-8}\right]$~Hz, while higher-frequency bins in this data set are dominated by noise. The NANOGrav $1\sigma$ and $2\sigma$ contours are given as the light and dark green regions in Fig.~\ref{fig5}. In the relevant frequency range, approximating the GWs from cosmic strings as a power law, we perform a logarithmic fit (represented by the red line in Fig.~\ref{fig5}) similar to that of Refs.~\cite{nanofit1,nanofit2,nanofit3} for $G\mu\sim G v^2_{\Phi}\sim G M_{DM}^2$ within the  DM mass range $M_{DM}\in \left[2\times 10^{14}~\text{GeV}, 2\times 10^{13}~\text{GeV}\right]$. Each point, marked a different shape, corresponds to a different benchmark value of the DM mass, as shown in the legenda. The interesting fact is that at $2\sigma$, the data constrains the SHDM mass to be within the range $3\times10^{13}~{\rm GeV} \lesssim M_{DM} \lesssim 10^{14}~{\rm GeV}$; this implies that the turning-point frequency is bound to be within $\sim 0.2-0.4$~Hz (from Eq.\ref{tevdm} and Eq.\ref{dmbr}).  Thus, whether the GWs possibly detected by NANOGrav come from the phase transition that gave a mass to the SHDM will be confirmed (or excluded) by the upcoming detectors LISA and DECIGO. Note also that such a spectral behaviour also distinguishes this scenario from those of Refs.~\cite{nanofit1,nanofit2,nanofit3} where at high-frequency the spectrum is flat ($\Omega_{GW}\sim f^0$).

\section{~Discussion and extensions }\label{sec5}
The predicted range of $f_*$ in this ${\rm PBH\rightarrow SHDM}$ mechanism could be obtained with a  high-scale $U(1)$ breaking and a well motivated early matter dominated phase. For instance, in Refs.~\cite{lepcs2,Borah:2022byb}, with a suitable choice of $G\mu$ and the decay width ($\Gamma_D$) of the long-lived field, the signal shown in Fig.\ref{fig4} (left) can be reproduced. Therefore to more robustly differentiate our scenario from models such as Refs.~\cite{lepcs2,Borah:2022byb}, we should look for additional signatures. In the simplest scenario, i.e., without going into the detail of a larger symmetry group such as GUT, one such signature could arise both from the DM and the PBH sectors of our model. Firstly, for certain interaction properties of DM, ultra-high-energy cosmic rays can play a crucial role to test or constrain the model (see appendix \ref{appe} for a constraint on the SHDM mass and life-time).   On the other hand, there could be further GWs from the PBH sector which could make the signatures of the model unique.  The most relevant example is GWs from the PBH density fluctuations that are very strong in amplitude for large $\beta$, which is a value that is preferred in our set-up as discussed in Sec.\ref{sec3}.\\

{\bf Gravitational waves from PBH density fluctuations}~~  In a PBH dominated universe, strong and sharply peaked gravitational waves  arise due to the inhomogeneous distribution of PBHs, as discussed recently in Ref.~\cite{bhgw5} and further developed in Refs.~\cite{bhgw6,Domenech:2021wkk}. The present-day energy density of such GWs is given by~\cite{bhgw6,Domenech:2021wkk}
\bea
\Omega_{GW}(t_0,f)^{I}\simeq \Omega_{GW}^{\rm peak}\left(\frac{f}{f_{\rm peak}}\right)^5\Theta
\left(f_{\rm peak}-f\right),\label{gu1}
\eea
where 
\bea
\Omega_{GW}^{\rm peak}\simeq 2\times 10^{-6} \left(\frac{\beta}{10^{-8}}\right)^{16/3}\left(\frac{M_{BH}}{10^7 \rm g}\right)^{34/9},~ f_{\rm peak}\simeq 1.7\times 10^3 {\rm Hz}\left(\frac{M_{BH}}{10^4 \rm g}\right)^{-5/6}.\label{gu2}
\eea
Two important points should be noted about Eq.\ref{gu2}. First, not to saturate the BBN bound on the effective number of neutrino species~\cite{bbn}, the initial PBH energy density  fraction parameter $\beta$ can not be arbitrarily large as shown in Fig.\ref{fig1}, where the solid blue line corresponds to $\beta_{\rm max}(M_{BH})$~\cite{bhgw6}.  Second,  for the PBH mass range $M_{BH}~\epsilon~\left[10^6{\rm g}-10^8{\rm g}\right]$ --- an allowed  window corresponding to SHDM production (see Fig.\ref{fig2}), the peak frequency $f_{\rm peak}$ lies within the testable range $f~\epsilon~\left[36{\rm Hz}\sim 0.7{\rm Hz}\right]$. Since, once the PBHs dominate, the number density of SHDM is independent of $\beta$, the prediction of the exact peak amplitude of $\Omega_{GW}(t_0,f)^{I}$ requires additional input. Nonetheless, as  the location of the turning point frequency $f_*$ is derived considering the fact that the entire DM relic density is in the form of SHDMs from PBHs, any other SHDM relics that may originate e.g., from gravitational production mechanisms~\cite{shdm1,shdm2} should be  diluted by considering large $\beta<\beta_{\rm max}$. Therefore, in this set-up, particularly for light SHDMs (e.g., $M_{DM}<10^{12}$ GeV), a spectral bump on the flat-plateau is strongly expected.  Conversely, the non-observation of such a peak would put further upper bounds on $N_{DM}^{i}$ in the $\rm PBH\rightarrow SHDM$ scenario. As a matter of fact, as discussed briefly in the next section, such peaked $\beta$-dependent GWs are not only  relevant to constrain $N_{DM}^i$, but also useful to test or constrain leptogenesis mechanisms along with SHDM in a more well-motivated model. 

\begin{figure}
\includegraphics[scale=.55]{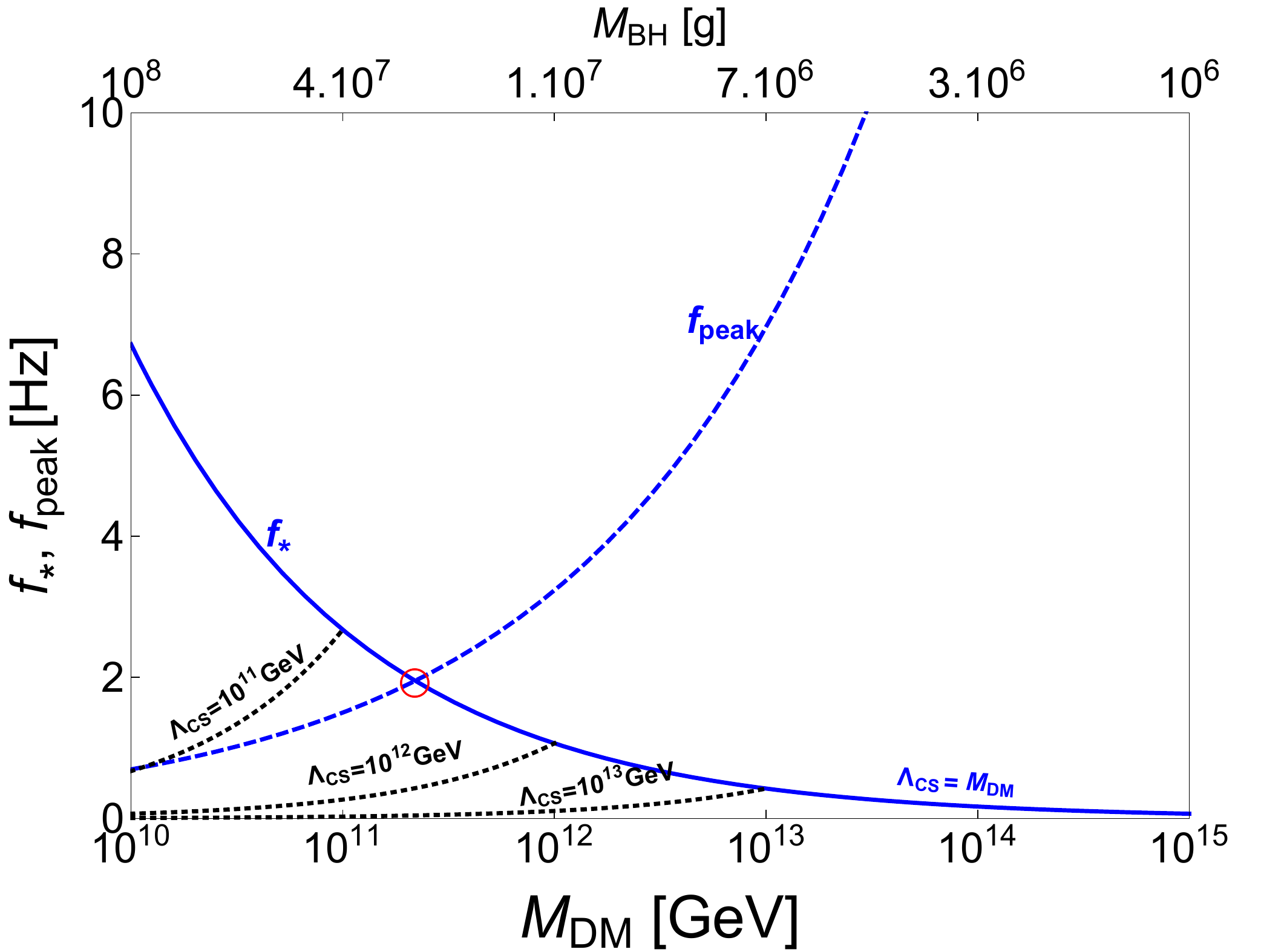} 
\caption{$M_{DM}$ vs. $f_{*,~{\rm peak}}$. The  blue solid line represents the turning point frequencies of the GWs from cosmic strings whereas the  blue dashed line represents the peak frequency of the GWs induced by PBH density fluctuation. The black dashed lines correspond to the cases where the  scale of symmetry breaking and the SHDM mass are different. The red circle marks the frequency $f=f_*=f_{\rm peak}$.} \label{fig6}
\end{figure}

Lastly, let us point out a remarkable coincidence that in this scenario there is  a significant overlap of the predicted  ranges of $f_*$ and $f_{\rm peak}$.  In general  $f_*$ and $f_{\rm peak}$ do not coincide and vary in opposite directions as $M_{DM}$ increases. This is shown in  Fig.\ref{fig6} with the blue curves. Interestingly,  at $M_{DM}\simeq 2.2\times 10^{11}$ GeV and $f\simeq 2$Hz, $f_*$ and $f_{\rm peak}$ coincide and the overall ($\Omega_{GW}^{CS}[f]+\Omega_{GW}^I[f]$) GW spectral shape becomes unique.\\

{\bf Leptogenesis}~~~Our scenario can be embedded in the well-motivated anomaly-free $U(1)_{B-L}$ extension of the SM (${\rm SM_{U(1)_{B-L}}}$)~\cite{ubl1,ubl2,ubl3} with three right handed (RH) neutrinos. A comprehensive summary of the appearance of $U(1)_{B-L}$ in GUT models can be found in~\cite{lepcs1}.  The inclusion of RH neutrinos in the SM solves the problem of non-vanishing neutrino mass via the seesaw mechanism~\cite{see1,see2}; this model can also explain the baryon asymmetry of the Universe via leptogenesis~\cite{leptor}.
 
The recent articles~\cite{nanofit3,lepcs1,lepcs2,lepcs3} consider the ${\rm SM_{U(1)_{B-L}}}$ model where all the three RH neutrinos decay quickly and show how the GWs from cosmic strings from the $U(1)$ symmetry breaking can be a probe of leptogenesis. However, this RH neutrino extension can be made more interesting by taking one of them to be rigorously stable due to an extra  parity symmetry or stable on cosmological scale because of its very feeble interactions with the SM: such RH neutrino then qualifies as a DM candidate. The model can still explain the neutrino oscillation data as well as baryogenesis via leptogenesis with the other two RH neutrinos~\cite{King:1999mb, Samanta:2019yeg}.  A number of works (e.g., see ~\cite{ublDM} and references therein) are dedicated to this possibility because a three-fermion SM extension solves three issues, namely, the neutrino masses, the baryon asymmetry of the Universe and the DM. 

Whereas the majority of these works consider freeze-out or freeze-in production mechanisms for light DM particles, in our set-up the RH neutrino DM has to be super-heavy. As shown in Ref.~\cite{romepbh}, if the RH neutrinos that produce the lepton/baryon  asymmetry are of PBH origin, then for a successful leptogenesis one finds an upper bound on the PBH mass that reads $M_{BH}\lesssim 13$~g; this is obviously not possible in our ${\rm PBHs\rightarrow SHDM}$ scenario (cf.\ Fig.~\ref{fig2}). Therefore, we need to consider thermal leptogeneis.  In the simplest scenario in which $\Lambda_{CS}=M_{DM}=M_R^{2}\simeq M_R^{1}$, where $M_R^{1,2}$ are the RH neutrino masses, leptogensis will be realised in the resonant regime~\cite{Pilaftsis:2003gt}. Defining the next-to-the-lightest RH neutrino mass $M_R^{2}=M_R^{1}\left(1+\delta\right)$ with $\delta \ll 1$, the  baryon asymmetry can be obtained as $N_{B-L} \simeq 6\times 10^{-17} \kappa \frac{\Lambda_{CS}}{\delta}$, where $\kappa\sim 10^{-2}$ is the efficiency of lepton asymmetry production~\cite{Pilaftsis:2003gt}.  Let us consider a benchmark value $\Lambda_{CS}=2.2\times 10^{11}$ as indicated with the red circle in Fig.\ref{fig6}.   With this choice, $N_{B-L}$ becomes $N_{B-L}\simeq 1.4\times 10^{-7}/\delta$. The final baryon asymmetry is then $N_{B-L}^f\simeq 1.4\times 10^{-7}\Delta\left(\beta\right)/\delta $, where $\Delta\left(\beta\right)$ is an entropy dilution factor that originates due to PBH domination. One has to compare $N_{B-L}^f$ with the observed value  $N_{B-L}^{Obs}\simeq 10^{-8}$~\cite{Planck:2015fie}. Since $\delta\ll 1$, the overproduced $N_{B-L}^f$ should be brought down to the observed one considering large $\beta$ and therefore  small $\Delta\left(\beta\right)$ (see Fig.\ref{fig3}). Recall that, as $\beta$ increases the peak amplitude of $\Omega_{GW}(t_0,f)^{I}$ also increases. Therefore, in this model there is an explicit relation between baryon asymmetry and  $\Omega_{GW}(t_0,f)^{I}$. From the observed baryon asymmetry, as the quasi-degeneracy between the RH neutrinos becomes strong ($\delta$ becomes smaller), the amplitude of $\Omega_{GW}(t_0,f)^{I}$ grows. A detailed discussion of leptogenesis and SHDM with $U(1)_{B-L}$ in a PBH dominated universe will be provided in our future publication.

\section{Summary and outlook}\label{sec6}

In this paper we consider a novel way to test the possibility that dark matter, specifically super-heavy dark matter with masses $M_{DM}\gtrsim 10^{10}$~GeV, is generated in the decay of primordial black holes $M_{BH}\lesssim 10^9$~g that evaporate before Big Bang Nucleosynthesis. We show how this possibility can be tested with gravitational waves, which have a strong amplitude because the dark matter is super-heavy. We explicitly build an example case in which the dark matter mass is generated by the breaking of a gauged $U(1)$ symmetry, which brings about a network of cosmic strings; it is these cosmic strings that are the sources of the strong gravitational waves that are within the sensitivity reach of current and planned gravitational waves detectors at nearly all frequencies.

This model possesses a further, unique feature that allows it to be distinguished from other scenarios.  The primordial black holes, in addition to acting as a very natural source of super-heavy dark matter, cause a break in the gravitational wave spectrum at higher frequencies wherein the spectral slope changes from $\Omega_{GW}\sim f^0$ to $\Omega_{GW}\sim f^{-1/3}$ beyond a turning-point frequency $f_*$. Given a measured amplitude ($\Omega_{GW}\sim M_{DM}$) at low frequencies, the model forecasts a spectral break $f_*\sim M_{DM}^{-2/5}$ at high frequency. Interestingly, if the dark matter mass is in the range $10^{11}$~GeV~$\lesssim M_{DM}\lesssim10^{15}$~GeV, the spectral break $f_*$ would be precisely within the sensitivity ranges of mid-band detectors such as BBO and DECIGO (and possibly LISA). {\it Thus the model exhibits a unique way to detect SHDM indirectly via the  low frequency-high frequency complementary measurements of GW spectral properties}.

We speculate that the recent finding of a stochastic common-spectrum process by NANOGrav and PPTA could be explained by the gravitational wave signal expected by a super-heavy dark matter with mass of about $3\times 10^{13}~\text{GeV} \lesssim M_{DM} \lesssim 10^{14}~\text{GeV}$. Espousing this interpretation, we predict that the gravitational wave spectrum exhibits a spectral break $f_*$ in the $\mathcal{O}(0.1)$~Hz range, and that this would be detectable by BBO and DECIGO. Therefore, whether dark matter is super-heavy and produced by the evaporation of primordial black holes would be tested by gravitational waves across nano-Hertz to Hertz frequencies, by probing both the low-frequency extended plateau and cross-checking this with the mid-frequency spectral break. We discuss how further GWs originating from the density fluctuations of PBHs makes our scenario even more unique.

As possible extensions of this work, we briefly discussed a simple embedding into a $U(1)_{B-L}$ model with three right-handed neutrinos, detail its prospects for also including a realistic explanation for the neutrino masses and baryogenesis. Furthermore, in this work we consider the effect of the PBHs on the GW spectrum from cosmic strings at the level of the background expansion of the Universe. However, the cosmic strings could potentially interact with the PBHs and form a black hole-string network that may lead to further spectral distortions in the GW spectrum~\cite{Vilenkin:2018zol}. A precise evolution of such a network would require numerical simulations, see~\cite{vilenkin}.

\section*{acknowledgement}
RS is supported by the MSCA-IF IV FZU - CZ.02.2.69/0.0/0.0/20 079/0017754 project and acknowledges European Structural and Investment Fund and the Czech Ministry of Education, Youth and Sports. FU is supported by the European Regional Development Fund (ESIF/ERDF) and the Czech Ministry of Education, Youth and Sports (MEYS) through Project CoGraDS - CZ.02.1.01/0.0/0.0/$15\_003$/0000437.

\appendix 
\section{Scaling loop number density and the plateau integral}\label{appa}

The  energy density parameter of gravitational waves at $t_0$  is expressed  as
\bea
\Omega_{GW}(t_0,f)=\frac{f}{\rho_c}\frac{d\rho_{GW}}{df}=\sum_k\Omega_{GW}^{(k)}(t_0,f),\label{feq}
\eea
where~\cite{cs9}
\bea
\frac{d\rho_{GW}^{(k)}}{df}=\int_{t_{osc}}^{t_0} \left[\frac{a(t)}{a(t_0)}\right]^4 P_{GW}(t,f_k)\frac{dF}{df}dt.\label{a2}
\eea
The quantity  $F=f \left[\frac{a(t_0)}{a(t)}\right]$ accounts for the red-shifting of the frequency and $P_{GW}(t,f_k)$ corresponds to the power emitted by the loops,  defined as~\cite{cs9}
\bea
P_{GW}(t,f_k)=G\mu^2\Gamma_k\int  n(l,t) \delta\left(f_k-\frac{2k}{l}\right) dl.\label{a3}
\eea
Integrating Eq.\ref{a3} over the loop lengths gives 
\bea
P_{GW}(t,f_k)=\frac{2kG\mu^2 \Gamma_k}{f_k^2} n(t,l_k)=\frac{2kG\mu^2 \Gamma_k}{f^2\left[\frac{a(t_0)}{a(t)}\right]^2}n\left(t,\frac{2k}{f}\left[\frac{a(t)}{a(t_0)}\right]\right).\label{a4}
\eea
From Eq.\ref{a4} and Eq.\ref{a2} one gets
\bea
\frac{d\rho_{GW}^{(k)}}{df}=\frac{2kG\mu^2 \Gamma_k}{f^2}\int_{t_{osc}}^{t_0} \left[\frac{a(t)}{a(t_0)}\right]^5 n\left(t,\frac{2k}{f}\left[\frac{a(t)}{a(t_0)}\right]\right)dt.
\eea
Therefore the energy density for the $k$-th mode is given by
\bea
\Omega_{GW}^{(k)}(t_0,f)=\frac{2kG\mu^2 \Gamma_k}{f\rho_c}\int_{t_{osc}}^{t_0} \left[\frac{a(t)}{a(t_0)}\right]^5 n\left(t,\frac{2k}{f}\left[\frac{a(t)}{a(t_0)}\right]\right)dt.\label{form1}
\eea

The loop number density $n\left(t,l_k=\frac{2k}{f}\left[\frac{a(t)}{a(t_0)}\right]\right)$ can be calculated with the analytical Velocity-dependent-One-Scale (VOS) model~\cite{vos1,vos2,vos3} that assumes the loop production function to be a delta function, i.e., all the loops are created with the same size. For a general equation of state parameter  $\omega$, the scaling number density $n_\omega\left(t,l_k\right)$ can be computed as~\cite{vos3,lepcs3}
\bea
n_\omega(t,l_{k}(t))=\frac{A_\beta}{\alpha}\frac{(\alpha+\Gamma G \mu)^{3(1-\beta)}}{\left[l_k(t)+\Gamma G \mu t\right]^{4-3\beta}t^{3\beta}},\label{genn0}
\eea
where $\beta=2/3(1+\omega)$. The parameter $A_\beta = 5.4~(\text{for}~w=1/3)$ or $A_\beta = 0.39~(\text{for}~\omega=0)$ is constant and represents the loop production efficiency in different cosmological epochs~\cite{vos3}. The constant values of $A_\beta$ are subjected to the assumption that the scaling regime is reached instantaneously when the network goes from one cosmological epoch to another~\cite{spb2}. \\

The analytic expression for the $\Omega_{GW}^{plt}$ can be derived by analysing the fundamental mode ($k=1$) and considering loop production as well as decay during radiation domination. A convenient way to obtain an expression for $\Omega_{GW}^{plt}$ is to convert the $t$ integral in Eq.\ref{gwf1} into a scale factor integral as 
\bea
\Omega_{GW}^{(1)}(t_0,f)=\frac{16\pi}{3\zeta(\delta)}\left(\frac{G\mu}{H_0}\right)^2\frac{\Gamma}{f a(t_0)}\int_{a_*}^{a_{eq}}H(a)^{-1} \left[\frac{a(t)}{a(t_0)}\right]^4 n\left(t,\frac{2}{f}\left[\frac{a(t)}{a(t_0)}\right]\right)da,\label{form3}
\eea
where
\bea
H=H_0\Omega_r^{1/2}\left(\frac{a(t)}{a(t_0)}\right)^{-2}~~{\rm with}~~\Omega_r\simeq 9\times10^{-5}
\eea
and the loop number density $n\left(t,l_1(t)\equiv \frac{2}{f}\left[\frac{a(t)}{a(t_0)}\right]\right)$ in Eq.\ref{genn0} (in radiation domination) is expressed as
\bea
n(t,l_{1}(t))=\frac{A_r}{\alpha}\frac{(\alpha+\Gamma G \mu)^{3/2}}{\left[\frac{2}{f}\left[\frac{a(t)}{a(t_0)}\right]+\Gamma G \mu/2H\right]^{5/2}(2H)^{-3/2}}.\label{numrad}
\eea

Let us mention also that the VOS model overestimates the number density of the loops approximately by an order of magnitude compared to the numerical simulations~\cite{cs9}. This is due to the fact that VOS model assumes that all the loops are of same size (a fraction of the horizon) at their production. In reality, there could be a distribution of $\alpha$. Numerical simulations find that only 10$\%$ of the energy from the long-string network goes to large loops ($\alpha\simeq 0.1$) while the remaining $90\%$~goes to highly boosted smaller loops which do not radiate GWs. Therefore, to be consistent with numerical simulations, we add the so-called calibration or normalisation factor $\mathcal{F}_\alpha\sim 0.1$ in Eq.\ref{genn0}~\cite{vos3}.

\section{Summing over all the modes}\label{appb}

The contributions from all the Fourier $k$ modes can easily be accounted for owing to the fact that
\bea
\Omega_{GW}(f)=\sum_k\Omega_{GW}^{(k)}(f)=\sum_k k^{-\delta}\Omega^{(1)}(f/k).\label{sumk}
\eea
Specifically, the spectral shape beyond the turning-point frequency $f_*$ changes in a significant way. For example, expanding the RHS of Eq.\ref{sumk} for the first few modes, i.e.,
\bea
\Omega_{GW}(f)&=&\sum_k k^{-\delta}\Omega^{(1)}(f/k) \nonumber \\ &=& 1^{-\delta}\Omega^{(1)}(f/1)+ m^{-\delta}\Omega^{(1)}(f/m)+ n^{-\delta}\Omega^{(1)}(f/n)+ r^{-\delta}\Omega^{(1)}(f/r)+...,
\eea
where the integers $m$, $n$ and $r$ obey $1<m<n<r$, we see that, should we keep on increasing the number of modes, there will be a critical point $k\equiv k_\Delta$ for which the amplitude $\Omega_{GW}^{(1)}(f_\Delta=f/k_\Delta)$ would contribute to the $\Omega_{GW}(f)$. Therefore, the sum can be performed in two parts:
\bea
\Omega_{GW}(f)&=&\sum_{k=1}^{k=k_{\Delta}}k^{-\delta}\Omega_{GW}^{(1)}(f/k>f_\Delta)+\sum_{k=k_{\Delta}}^{k=k_{max}}k^{-\delta}\Omega_{GW}^{(1)}(f/k<f_\Delta)\\
&=&\sum_{k=1}^{k=k_{\Delta}}k^{-\delta}\Omega_{GW}^{\rm plt}\left(\frac{f_\Delta}{f/k}\right)+\sum_{k=k_{\Delta}}^{k=k_{max}}k^{-\delta}\Omega_{GW}^{\rm plt}\label{2sum}
\eea
That gives
\bea
\Omega_{GW}(f)\simeq \Omega_{GW}^{\rm plt} \left( \frac{f_\Delta}{f}\right)^{\delta-1}~~{\rm i.e.,}~~\Omega_{GW}(f)\propto\begin{cases}f^{-1/3} ~~{\rm cusps}\\f^{-2/3}~~{\rm kinks}. \end{cases}
\eea
upon using the asymptotic expansion of the Euler-Maclaurin series for the first term and the expansion of the Hurwitz zeta function for the second term.

\section{PBH factsheet}\label{appc}

{\it Life-time, evaporation and condition for PBH domination}: For the radiation energy density to dominate at the time of PBH evaporation $t_{ev}$, one has 
\bea
r(t_{ev})\equiv\frac{\rho_{ BH}(t_{ev})}{\rho_{\rm R}(t_{ev})}<1.\label{rad dom}
\eea
The ratio of the $r$ parameters at the time of black hole formation and evaporation is therefore
\bea
\frac{r(t_{ev})}{r(t_{Bf})}=\frac{a(t_{ev})}{a(t_{Bf})}=\left(\frac{t_{ev}}{t_{Bf}}\right)^{1/2}.\label{rr}
\eea
Using the first Friedmann equation 
\bea
H(t)^2=\frac{8\pi}{3M_{Pl}^2}\rho_{R}(t),~~ {\rm with}~~M_{Pl}=1.22\times 10^{19}~{\rm GeV}\label{frd1}
\eea
along with the expression for the Hubble parameter $H(t)$ in radiation domination and the radiation energy density 
\bea
H=\frac{1}{2t}, ~~\rho_{R}(T)=\frac{\pi^2 g_*(T) T^4}{30},\label{radeng}
\eea
where $g_*(T)$ ($\simeq$ 106.75 in SM) is the effective degrees of freedom that contribute to the radiation.  Eq.\ref{rr} can be rewritten as
\bea
\frac{r(t_{ev})}{r(t_{Bf})}=\left(\frac{g_*(T_{Bf})}{g_*(T_{ev})}\right)^{1/4}\frac{T_{Bf}}{T_{ev}}.\label{key}
\eea
Eq.\ref{rad dom} now translates to the condition
\bea
r(t_{Bf})\equiv \beta <\left(\frac{g_*(T_{ev})}{g_*(T_{Bf})}\right)^{1/4}\frac{T_{ev}}{T_{Bf}}.\label{condrad}
\eea
The above condition on $\beta$ can be expressed fully in terms a single free parameter $M_{BH}$---we work in the approximation that the PBH mass spectrum is monochromatic. Assuming radiation domination at black hole formation, the mass of a PBH originating from gravitational collapse is approximately given by the energy density enclosed in a post-inflationary particle horizon~\cite{pdmrev}, i.e.,
\bea
M_{BH}=\gamma\frac{4}{3}\pi(H_{Bf}^{-1})^3 \rho_{Bf}~~{\rm with}~~\rho_{Bf}=\frac{3 H_{Bf}^2 M_{Pl}^2}{8\pi}, ~~H_{Bf}=\frac{1}{2t_{Bf}}.\label{ph}
\eea
The quantity $\gamma\simeq 0.2$ depends on the details of the gravitational collapse mechanism. From Eq.\ref{ph}, the PBH formation time $t_{Bf}$ is calculated as
\bea
t_{Bf}=\frac{M_{BH}}{M_{Pl}^2\gamma}.
\eea
Now using Eq.\ref{frd1} and Eq.\ref{radeng} one obtains the PBH formation temperature as
\bea
T_{Bf}=\left(\frac{45 \gamma^2}{16\pi^3 g_*(T_{Bf})}\right)^{1/4}\left(\frac{M_{Pl}}{M_{BH}}\right)^{1/2}M_{Pl}.\label{tbf}
\eea

Let us, for now, assume that radiation dominates throughout the evolution of the PBH---we will discuss the case in which PBHs come to dominate for a period of time in Appendix~\ref{appd}. The Hubble parameter is given by
\bea
H(t_{ev })^2=\frac{1}{4t_{ev}^2}\simeq \frac{1}{4\tau^2},
\eea
where the $\tau$ is the lifetime of the PBH. Therefore, again, using  Eq.\ref{frd1} and  Eq.\ref{radeng}, $T_{ev}$ can be obtained as
\bea
T_{ev}=\left(\frac{45 M_{Pl}^2}{16\pi^3 g_*(T_{ev})\tau^2}\right)^{1/4}.\label{teva}
\eea
The lifetime of the PBH is obtained from the dynamics of the mass loss of a PBH via Hawking evaporation~\cite{hr}. The rate at which PBH loses mass is given by
\bea
-\frac{dM_{BH}}{dt}=f_{ev}(4\pi r_{BH}^2)\frac{dE}{dt},\label{hwkrad}
\eea
where the $\frac{dE}{dt}$ is obtained as
\bea
\frac{dE}{dt}=2g_{BH}(T_{BH})\pi^2\int_0^\infty d\nu\frac{\nu^3}{{\rm exp} (2\pi\nu/T)-1}=\frac{\pi^2}{120}g_{BH}(T_{BH}) T_{BH}^4.
\eea
The quantity $f_{ev}$ is the efficiency of PBH evaporation and $r_{BH}$ is the Schwarzschild radius. The quantity $g_{BH}(T_{BH})$ counts the bosonic and fermionic degrees of freedom for $T<T_{BH}$.  Recalling the PBH temperature ($T_{BH}=M_{Pl}^2/8\pi M_{BH}$~\cite{hr}) and using  $r_{BH}=2GM_{BH}$, Eq.\ref{hwkrad} is recast as 
\bea
\frac{dM_{BH}}{dt}=-\frac{\mathcal{G} g_{*B}(T_{BH})}{30720\pi}\frac{M_{Pl}^4}{M^2_{BH}},
\eea
where one uses $f_{ev}g_{BH}(T_{BH})=\mathcal{G}g_{*B}(T_{BH})$. The  lifetime $\tau$ is then obtained as
\bea
\tau=\int_{t_{Bf}}^{t_{ev}}dt=-\int_{M_{BH}}^0d M_{BH}\frac{30720\pi M_{BH}^2}{\mathcal{G}g_{*B}(T_{BH})M_{Pl}^4}=\frac{10240\pi M_{BH}^3}{\mathcal{G}g_{*B}(T_{BH})M_{Pl}^4}.\label{lt}
\eea
Now combining Eq.\ref{lt}, Eq.\ref{teva} and Eq.\ref{tbf}, we can recast Eq.\ref{condrad} as 
\bea
\beta<\gamma^{-1/2}\left(\frac{\mathcal{G }g_{*B}(T_{BH})}{10240\pi}\right)^{1/2}\frac{M_{Pl}}{M_{BH}}.\label{beta_bound}
\eea

{\it The Friedmann equations:}
\bea
\frac{d\rho_R}{dt}+4H\rho_R=-\frac{\dot{M}_{BH}}{M_{BH}}\rho_{BH},\label{be1}\\
\frac{d\rho_{BH}}{dt}+3H\rho_{BH}=+\frac{\dot{M}_{BH}}{M_{BH}}\rho_{BH},\label{be2}\\
\frac{ds}{dt}+3Hs=-\frac{\dot{M}_{BH}}{M_{BH}}\frac{\rho_{BH}}{T},\label{be3}
\eea
where the Eq.\ref{be3} represents the non-conservation of entropy $\tilde{S}\sim s a^3$ due to PBH evaporation.

\section{DM and PBH number densities}\label{appd}

In a simplified approach (neglecting the momentum distribution of the emitted particles), the differential number of a particle species $`X$' emitted by a black hole can be computed as~\cite{br0a}
\bea
d\bar{n}=dE/3T_{BH}=\frac{M_{Pl}^2}{24 \pi} \frac{1}{T_{BH}^3}dT_{BH},
\eea
where we have used 
\bea
dE\equiv -d(M_{BH})=\frac{M_{Pl}^2}{8\pi}\frac{dT_{BH}}{T_{BH}^2}
\eea
and the mean energy of the radiated particles is $\bar{E}=3T$. The total number of $X$ particles emitted in the case of a complete evaporation is given by
\bea
\bar{n}_X=\frac{g_X}{g_{*B}}\int_{T_{BH}}^\infty d\bar{n}=\frac{4\pi}{3}\frac{g_X}{g_{*B}}\left(\frac{M_{BH}}{M_{Pl}}\right)^2 ~~{\rm for}~~T_{BH}>M_X,\label{pp1}\\
\bar{n}_X=\frac{g_X}{g_{*B}}\int_{M_X}^\infty d\bar{n}=\frac{1}{48\pi}\frac{g_X}{g_{*B}}\left(\frac{M_{Pl}}{M_X}\right)^2~~{\rm for}~~T_{BH}<M_X, \label{pp2}
\eea
where $M_X$ and $g_X$ are the mass and internal degrees of freedom of the radiated particles. The quantity $N_{\rm BH}^{\rm ev}$ is obtained as
\bea
N_{\rm BH}^{\rm ev}=\left(\frac{n_{\rm BH}}{n_{f,\rm eq}^{\rm ur}} \right)_{T_{ev}}=\frac{M_{pl}^2}{6\pi \tau^2 M_{BH}}\left( \frac{g_f T_{ev}^3}{\pi^2}\right)^{-1},\label{NBHeva0}
\eea
where we have used $n_{\rm BH}=\rho_{\rm BH}/M_{BH}$ with $\rho_{\rm BH}$ calculated from the Friedmann equation when PBHs dominate: $H\simeq 2/3\tau$. Proceeding in the same way as in Eq.\ref{teva}, the evaporation temperature in the case of PBH domination can be calculated as
\bea
T_{ev}=\left(\frac{5 M_{Pl}^2}{\pi^3 g_*(T_{ev})\tau^2}\right)^{1/4}.\label{tevab}
\eea
Combining Eq.\ref{NBHeva0} and Eq.\ref{tevab} we find
\bea
N_{\rm BH}^{\rm ev}=\left(\frac{\pi}{6g_f}\right)\left(\frac{\pi^3g_*(T_{ev})}{5}\right)^{3/4}\left(\frac{\mathcal{G}g_{*B}(T_{BH})}{10240\pi}\right)^{1/2}\left(\frac{M_{Pl}}{M_{BH}}\right)^{5/2}.\label{NBHeva1}
\eea
\section{UHECR constraints}\label{appe}
If the DM is indeed super-heavy and it is not exactly stable, it is possible to test and constrain its properties using UHECRs data, see, e.g., \cite{Marzola:2016hyt,Kalashev:2016cre} and references therein.  Indeed, when SHDM particles decay, they initiate a cascade which produces a slew of very high energy photons, neutrinos, and protons, which can be detected in UHECR experimental facilities.  The most stringent constraints come from the non-detection of ultra-high-energy photons, as they are more numerous than protons and much easier to detect than neutrinos.  For the range of masses we are interested in, the most significant result comes from the non-observation of any UHECR, and therefore also photons, at primary energies above \(E = 10^{11.3}~\text{GeV}\) at the Pierre Auger Observatory~\cite{Alcantara:2019sco}, which translates in an upper limit on the integral flux above this energy of
\begin{align}
    \Phi_\text{exp}(E>10^{11.3}~\text{GeV}) \lesssim \frac{3.6\cdot10^{-5}}{\text{km}^2\,\text{sr}\,\text{yr}} \,. \label{phi_exp}
\end{align}
At lower energies the limits from the Pierre Auger Observatory and the Telescope Array collaboration are similar and approximately two orders of magnitude lower~\cite{PierreAuger:2021mjh,TelescopeArray:2021fpj}.

The predicted flux of photons from SHDM decay instead can be written as~\cite{Marzola:2016hyt,Kalashev:2016cre,Chianese:2021jke}
\begin{align}
    J(E) = \frac{1}{4\pi M_{DM} \tau_{DM}} \frac{\dd N_\gamma}{\dd E} \int \dd s \, \rho_{DM}(s) \,, \label{j_thy}
\end{align}
where the integral of the SHDM density \(\rho_{DM}(s)\) over the line of sight \(s\) can be written as an integral over Galactic radius \(r\) as
\begin{align}
    \int \dd s \, \rho_{DM}(s) = \left[ 2\int_{r_\odot\sin\theta}^{r_\odot} + \int_{r_\odot}^{r_H} \right] \dd r \, r \, \frac{\rho_{DM}(r)}{\sqrt{r^2-r_\odot^2\sin^2\theta}} \,, \label{int_rho}
\end{align}
with the SHDM density itself---we adopt the Einasto profile here---being given by
\begin{align}
    \rho_{DM}(r) \equiv \rho_s \exp\left\{-\frac{2}{\alpha}\left[\left(\frac{r}{r_s}\right)^\alpha-1\right]\right\} \,. \label{rho_DM}
\end{align}
In these expressions \(\rho_s = 0.033~\text{GeV}/\text{cm}^3\), \(r_s = 28.44~\text{kpc}\), \(\alpha = 0.11\), \(r_H = 260~\text{kpc}\) and \(r_\odot = 8.33~\text{kpc}\).  The SHDM density integral Eq.\ref{int_rho} depends on the angle \(\theta\) between the line of sight and the Sun-Galactic Centre axis; in our numerical estimates we will average the photon flux over the whole sky for simplicity.

The flux Eq.\ref{j_thy} depends on the SHDM lifetime \(\tau_{DM}\) and mass \(M_{DM}\), as well as on the differential number of photons produced by each SHDM decay
\begin{align}
    \frac{\dd N_\gamma}{\dd E} \equiv \frac{2}{M_{DM}} \frac{\dd N_\gamma}{\dd x} \equiv \frac{2\hat{N}_\gamma}{M_{DM}} \left(\frac{x}{\hat{x}}\right)^{-\gamma} \,,
\end{align}
with \(\gamma = 1.9\), \(\hat{N}_\gamma \approx 10^8\) and \(\hat{x} = 10^{-5}\).  This simplification is within 10\% (or better) of the true numerical result provided the lower limit of the integral flux satisfies \(x \equiv 2E/M_{DM} \lesssim0.1\)---this is certainly true for the UHECR energies and SHDM masses we are considering in this work.

Integrating the flux Eq.\ref{j_thy} above the energy cut \(E_\text{cut} = 10^{11.3}~\text{GeV}\) gives us the predicted integral flux \(\Phi_\gamma\) which needs to satisfy the constraint Eq.\ref{phi_exp}.  For the SHDM that are of most interest in this work we therefore obtain
\begin{align}
    \tau_{DM} \gtrsim 10^{24} \left(\frac{M_{DM}}{\text{GeV}}\right)^{-0.1}~\text{yr} \,. \label{tau_DM_bound}
\end{align}
For lower values of the mass \(M_{DM}\in[10^{10},10^{12}]~\text{GeV}\) the limit reads \(\tau_{DM} \gtrsim 10^{21}~\text{yr}\).

{}
\end{document}